\documentclass[apj]{emulateapj}
\usepackage{natbib}
\usepackage{epstopdf}
\usepackage{CJK}
\usepackage{color}
\usepackage{pshan}
\usepackage{amsmath}
\usepackage{diagbox}

\newcommand{\beq}{\begin{equation}}
\newcommand{\eeq}{\end{equation}}
\newcommand{\bea}{\begin{eqnarray}}
\newcommand{\eea}{\end{eqnarray}}

\begin{document}
\begin{CJK}{UTF8}{mj}
\title{Large-scale variation in reionization history caused by baryon-dark matter streaming velocity}

\author{Hyunbae Park (박현배)\altaffilmark{1}}
\author{Paul R. Shapiro\altaffilmark{2}}
\author{Kyungjin Ahn\altaffilmark{3}}
\author{Naoki Yoshida\altaffilmark{1,4,5}}
\author{Shingo Hirano\altaffilmark{6}}
\altaffiltext{1}{Kavli IPMU (WPI), UTIAS, The University of Tokyo, Kashiwa, Chiba 277-8583, Japan; hyunbae.park@ipmu.jp}
\altaffiltext{2}{Texas Cosmology Center and the Department of Astronomy, The University of Texas at Austin, 1 University Station, C1400, Austin, TX 78712, USA; shapiro@astro.as.utexas.edu}
\altaffiltext{3}{Department of Earth Sciences, Chosun University, Gwangju 61452, Republic of Korea; kjahn@chosun.ac.kr}
\altaffiltext{4}{Department of Physics, School of Science, The University of Tokyo, Bunkyo, Tokyo 113-0033, Japan}
\altaffiltext{5}{Research Center for the Early Universe, School of Science, The University of Tokyo, Bunkyo, Tokyo 113-0033, Japan}
\altaffiltext{6}{Department of Earth and Planetary Sciences, Faculty of Sciences, Kyushu University, Fukuoka, Fukuoka 819-0395, Japan}

\begin{abstract} 
At cosmic recombination, there was supersonic relative motion between baryons and dark matter, which originated from the baryonic acoustic oscillations in the early universe. This motion has been considered to have a negligible impact on the late stage of cosmic reionization because the relative velocity quickly decreases. However, recent studies have suggested that the recombination in gas clouds smaller than the local Jeans mass ($\lesssim$ $10^8~M_\sun$) can affect the reionization history by boosting the number of ultraviolet photons required for ionizing the intergalactic medium. 
Motivated by this, we performed a series of radiation-hydrodynamic simulations to investigate whether the streaming motion can generate variation in the local reionization history by smoothing out clumpy small-scale structures and lowering the ionizing photon budget.  We found that the streaming velocity can add a variation of $\Delta z_e$ $\sim$ $0.05$ $-$ $0.5$ in the end-of-reionization redshift, depending on the level of X-ray preheating and the time evolution of ionizing sources. The variation tends to be larger when the ionizing efficiency of galaxies decreases toward later times. Given the long spatial fluctuation scales of the streaming motion ($\gtrsim 100$ Mpc), it can help to explain the Ly$\alpha$ opacity variation observed from quasars and leave large-scale imprints on the ionization field of the intergalactic medium during the reionization. The pre-reionization heating by X-ray sources is another critical factor that can suppress small-scale gas clumping and can diminish the variation in $z_e$ introduced by the streaming motion. 
\end{abstract}

\section{Introduction}

The Epoch of Reionization (EOR) is an exciting test site for theoretical models of structure formation. The 
evolution of the global ionization fraction of the intergalactic medium (IGM) is driven by competition between the production of the ionizing photons from galaxies and the consumption of the photons by the intergalactic gas. Observational data from this era has accumulated over the last two decades, 
which have enabled us to constrain the luminosity function of the early galaxies  \citep[e.g.,][]{2013MNRAS.432.2696M,2015ApJ...810...71F,2015ApJ...803...34B}.

The first direct constraint on the reionization history came from the Gunn$-$Peterson (GP) trough feature in quasar spectra, where the Ly$\alpha$ opacity of the IGM rises steeply at $z\sim 6$ marking the end of reionization \citep{2004AJ....127.2598S,2006ARA&A..44..415F}. Measurements of the anisotropies of the cosmic microwave background (CMB) provide constraints on the integrated Thompson optical depth of the IGM. The latest results from the Planck Collaboration imply that reionization was in the midway at $z=7.7\pm0.7$ \citep{2018arXiv180706209P}. Also, the observed decline in the number density of Lyman-$\alpha$ emitters \citep{2010ApJ...725L.205F,2011ApJ...743..132P,2012ApJ...744..179S, 2016ApJ...818...38S, 2018PASJ...70S..16K,2018ApJ...864..103J,2018ApJ...856....2M,2020ApJ...904..144J} and detailed analysis of the spectra of high-redshift quasars \citep{2011Natur.474..616M, 2018Natur.553..473B,2018ApJ...864..142D} constrain the IGM neutral fraction, which provides indirect information on the nature of the sources of reionization \citep[e.g.,][]{2013ApJ...768...71R,2015ApJ...802L..19R,2019ApJ...879...36F}.

The main interest of the present study is to explain the apparent variation in the Ly$\alpha$ opacity repeatedly reported by observations of distant quasars \citep{2006ARA&A..44..415F,2008MNRAS.386..359G,2011MNRAS.416L..70B,2015MNRAS.447.3402B}. While it has been often regarded that the reionization ended before $z=6$, some sightlines show a GP trough stretching well below that redshift, suggesting a substantial variation in the Ly$\alpha$ opacity at $\sim50~\rm{Mpc}$ scales. The most extreme case found is a $\sim150~\rm{Mpc}$ long trough stretching from $z\approx6$ to 5.5 in the sightline of quasar ULAS J0148+0600 reported by \citet{2015MNRAS.447.3402B}. It is not understood how this quasar could coexist with other quasars that do not show GP troughs up to $z=6$ because it is difficult to explain the large-scale fluctuations in the ultraviolet (UV) background and IGM temperature inferred from the observations.

The attempts to explain the Ly$\alpha$ opacity variation generally fall into two categories. One is to assume that reionization completed before $z=6$ and the opacity variation comes from the variations in the IGM temperature \citep{2015ApJ...813L..38D} or in the UV background \citep{2016MNRAS.460.1328D,2017MNRAS.468.4691D,2017MNRAS.465.3429C,2018MNRAS.473..560D}. The other is to relax the assumption that reionization finished by $z=6$ and consider the possibility that reionization ended later in void regions with low galaxy density \citep{2019MNRAS.485L..24K,2020MNRAS.491.1736K,2020MNRAS.494.3080N}. Interestingly, \cite{2020MNRAS.494.3080N} have shown that both approaches could reproduce the large-scale opacity variation under certain conditions. \cite{2019MNRAS.487.1047M} and \cite{2020MNRAS.499.1640M} suggested that the variation in the IGM temperature leaves a measurable imprint on the Ly$\alpha$ forest at $z\sim2$, which will be measured by the Dark Energy Spectroscopic Instrument survey.

 \begin{figure*}
\begin{center}
\includegraphics[scale=0.6]{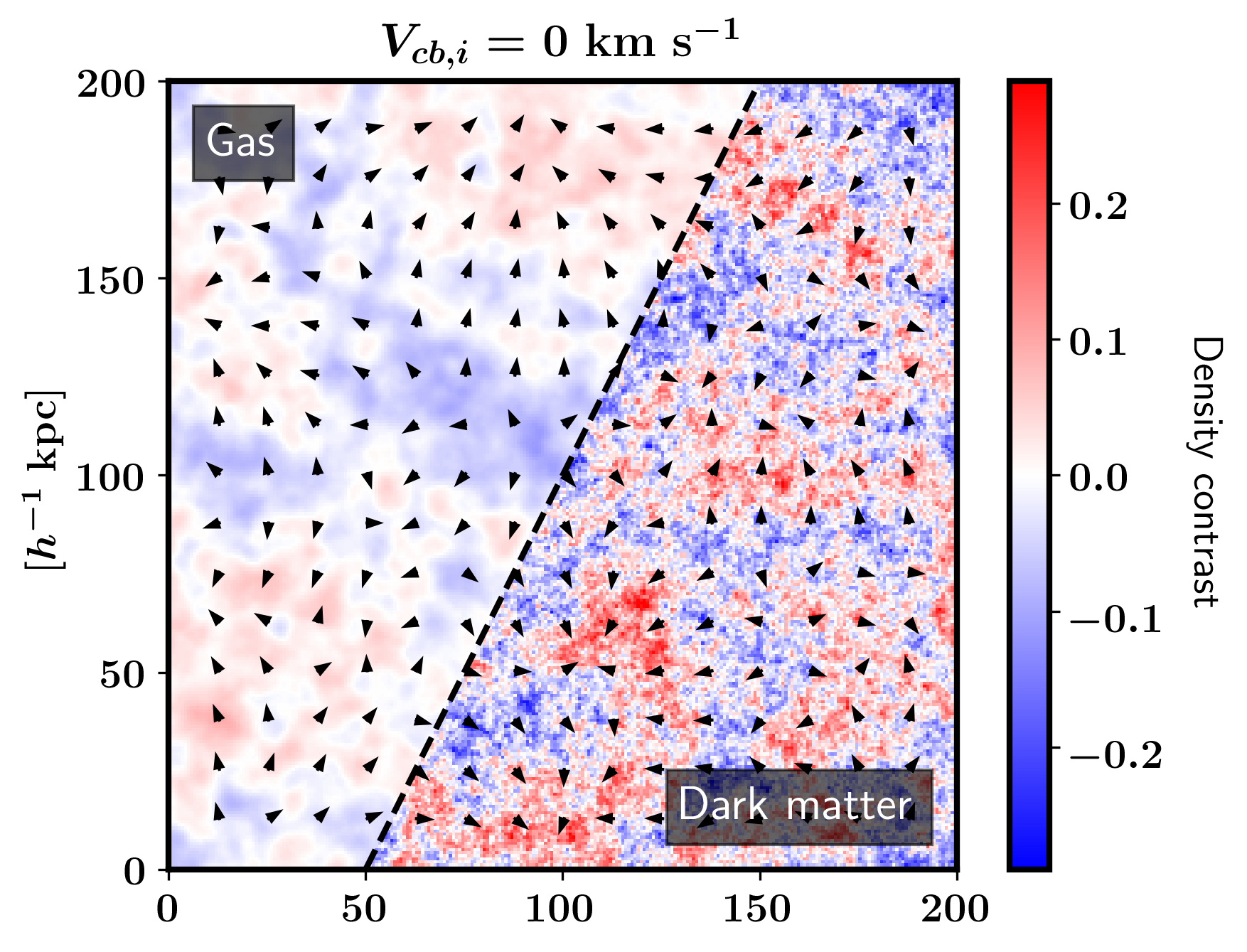}
\includegraphics[scale=0.6]{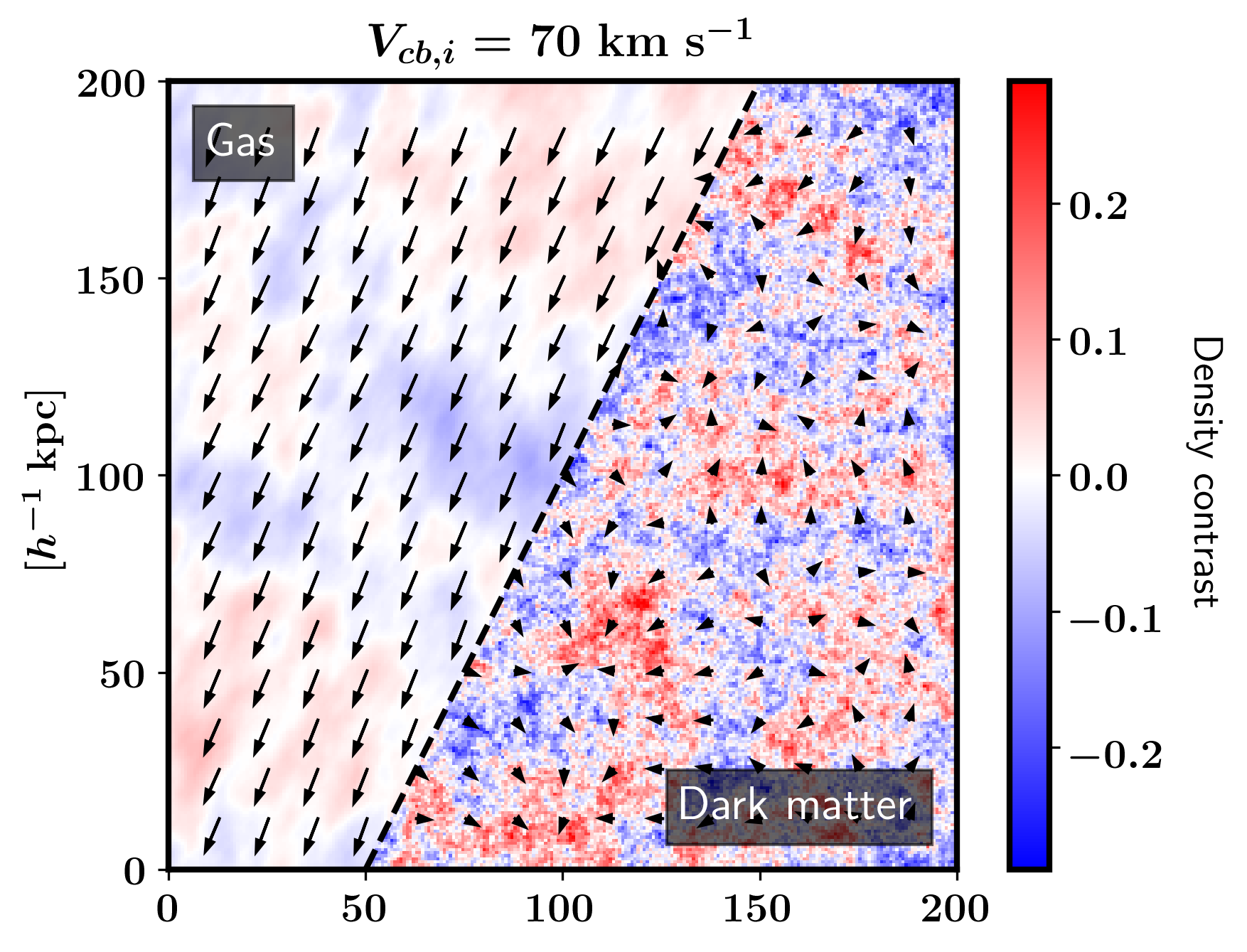}
\caption{Initial density and velocity fields of the simulations with zero streaming velocity (left panel) and a streaming velocity of $V_{cb,i}=70~{\rm km}~{\rm s}^{-1}$ (right panel). The upper left and lower right parts of each panel describe the gas and dark matter components. The color map and the vectors depict the spatial distribution of density and velocity, respectively.}
\label{fig:ICmap}
\end{center}
\end{figure*}

In this study, we explore the possibility that the relative streaming motion between baryons and dark matter causes the observed Ly$\alpha$ opacity variation by generating large-scale inhomogeneity in reionization. The streaming motion is a remnant of the baryonic acoustic oscillation (BAO), which froze out at the cosmic recombination with its amplitude around $\sim 30~{\rm km}~{\rm s}^{-1}$ ranging from $0-70~{\rm km}~{\rm s}^{-1}$ \citep{2010PhRvD..82h3520T}. It is known to hinder the formation of the Population III stars in minihalos (MHs) at the Cosmic Dawn ($z\sim 20 - 30$) \citep{2011ApJ...736..147G,2012ApJ...760....4O,2013ApJ...763...27N,2013ApJ...771...81R,2016PhRvD..93b3518A,2019MNRAS.484.3510S} and generate large-scale imprints on the 21cm fluctuations from the epoch \citep{2012ApJ...760....3M,2012Natur.487...70V,2019PhRvD.100f3538M}. However, its impact on the EOR ($z\sim6$) has been considered unimportant because the streaming motion does not affect the galaxies in atomically cooling halos with $M>10^8 M_\odot$, which produce most of the ionizing photons during the reionization era \citep{2011ApJ...730L...1S,2014MNRAS.437L..36F}.


Here, we pay attention to the photon consumption in structures below the Jeans mass of photoionized gas ($\sim 10^8~ M_\odot$). The recombination rate of photoionized gas during reionization -- as sometimes expressed by a clumping factor by which the average recombination rate is enhanced to account for unresolved density inhomogeneity -- has long been suggested to depend upon the inclusion of inhomogeneity on the smallest scales between the Jeans-filtered scale of the photoionized gas following its reionization and the Jeans-filtered scale of the pre-reionization gas \citep{2003AIPC..666...89S,2004MNRAS.348..753S}. This has been demonstrated in a series of papers, starting from the first rad-hydro simulations of the photoevaporation of MHs by the intergalactic I-fronts that propagate during the EOR \citep{2004MNRAS.348..753S,2005MNRAS.361..405I} and the application of those results to model the impact of this process on the progress of reionization \citep{2005ApJ...624..491I}.  

Meanwhile, the clumping factor and recombination rate enhancement associated with this small-scale structure in a three-dimensional cosmological density field, in the filamentary structure outside of MHs, as well, was subsequently described by
\citep{2013ApJ...763..146E} in an approximate way, by taking a snapshot of a tiny, highly resolved cosmological $N$-body + hydro simulation without radiative transfer or ionizing radiation, and then post-processing the static gas density field by introducing ionizing radiation, and relaxing the outcome to solve for the ``Str\"{o}mgren surface" separating the ionized exteriors of halos from the self-shielded interiors of halos. This work established the demands for spatial and mass resolution required to make the simulated value of the total recombination rate in the ionized regions converge.   

Following that static calculation, \citet[][hereafter P16]{2016ApJ...831...86P} developed the Gadget-RT code to perform this calculation with fully coupled radiation hydrodynamics. The results showed that the impact of a small-scale gas structure (SSGS) on the global progress of reionization could be significantly affected by the additional recombinations that result from these small-scale density fluctuations, including both the MHs and the fluctuating IGM around them, and parameterized the effect for application to future models. Recently, this problem has been revisited by \citet[][hereafter D20]{2020ApJ...898..149D}, who confirmed the importance of the SSGS contribution in the ionization photon budget of reionization and showed a quantitative impact on the global reionization history. This small-scale recombination can be implemented in large-scale reionization simulations as a sub-grid correction, and the predictions for reionization observable are affected significantly by the correction \citep{2020MNRAS.491.1600M}.

In the meantime, \cite{2018MNRAS.474.2173H} showed that the streaming motion disrupts a large portion of SSGSs at $z\sim6$  despite its characteristic decay over time. Thus, our purpose here is to explore the impact of this effect on the clumping factor and recombination rate boost reported by P16 and D20. In particular, we shall focus our interest here on the possibility that the effect is modulated on large scales, correlated with the BAOs.

Another potentially significant effect is the pre-reionization heating by X-ray background. During the reionization, X-rays from accreting black holes are expected to have heated the IGM to some level \citep{2004MNRAS.352..547R,2018MNRAS.476.1174E}. 
The formation of SSGSs in the pre-reionization IGM os largely owed to its low gas temperature. Thus, intense X-ray preheating would likely suppress SSGSs. The exact value of the gas temperature in the pre-reionization phase is highly uncertain. Thus, we shall consider it as a free parameter between $10$ and $1000$ K in this work. 

Therefore, our goal is to assess the impact of the streaming motion and the X-ray preheating on the recombination by the SSGSs and the resulting reionization history using a series of hydrodynamic simulations developed by P16. In a similar work, \citet[][hereafter C20]{2020ApJ...898..168C} investigated the signature of the streaming motion in the 21 cm signal. This study focuses on the impacts on the reionization history. 

The rest of the paper is organized as follows. In Section~\ref{sec:sim}, we explain how we initialize our simulations with and without the streaming motion. In Section~\ref{sec:PRIGM}, we explain how we evolve the initial conditions (ICs) of the IGM until it is exposed to the ionizing radiation. In Section~\ref{sec:IGMaR}, we report the simulated results on the clumpiness of the IGM after being exposed to ionizing background radiation. In Section~\ref{sec:RHist}, we calculate the reionization history from the clumping factor measured from our simulations to assess the impact of the streaming velocity and X-ray preheating on reionization. In Section~\ref{sec:discussion}, we summarize and discuss our results. For the rest of this paper, we assume a Lambda cold dark matter cosmology consistent with the Wilkinson Microwave Anisotropy Probe 9 yr results \citep{2013ApJS..208...19H}: $\Omega_{m,0}=0.276$, $\Omega_{b,0}=0.045$, $h=0.703$, $\sigma_8$=0.8 and $n_s=0.961$.

\section{IC\MakeLowercase{s}} \label{sec:sim}

We create cosmological ICs at $z=200$ with $256^3$ particles for gas and dark matter each in a 200 $h^{-1}$ kpc box. In this setup, the gas and dark matter particle masses are $44$ and $8.4~h^{-1}~M_\odot$, respectively. P16 showed that this mass resolution well captures the internal gas structure of MHs during the evaporation. In a similar work, D20 also arrived at a similar conclusion in their convergence test (see appendix C of their work).

The simulation box size ($200~h^{-1}~{\rm kpc}$) in this study is not large enough to capture the large-scale variation in density. The local overdensity of such a small volume can deviate substantially from zero, and P16 and D20 showed that the recombination rate depends sensitively on the local overdensity. The box size and resolution test in Appendix C of C20 showed that our box size would miss large-scale contributions to the clumping factor, but they are not affected much by the streaming motion. We limit the scope of this study to reporting the qualitative impacts of the streaming motion on the recombination in a mean-density volume.

 \begin{figure*}
\begin{center}
\includegraphics[scale=1.2]{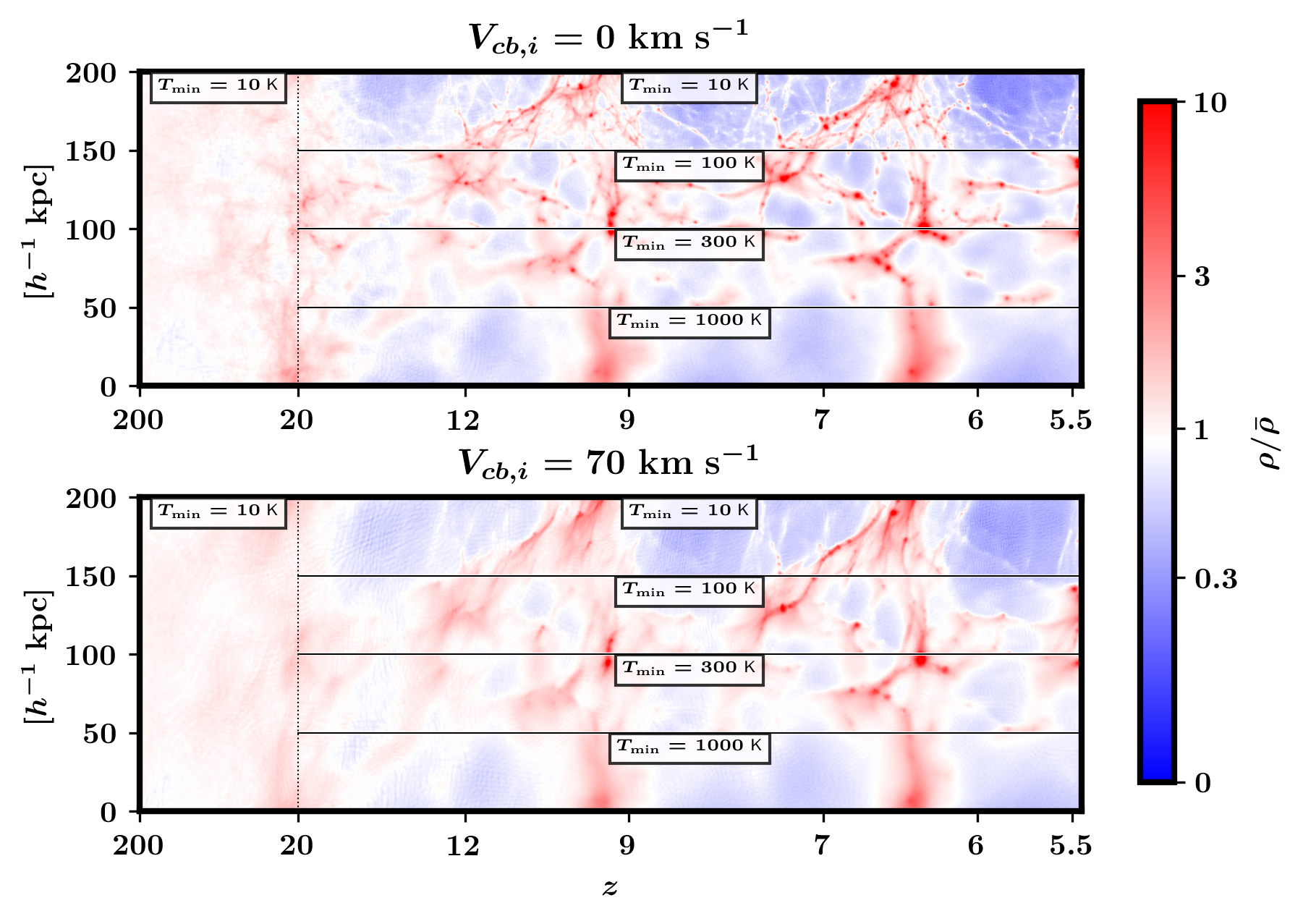}
\caption{Projected density field of the pre-reionization IGM evolving from $z=200$-$5.5$. We stitch vertical strips from different snapshots to depict the time evolution of the density field along the $x$-axis. The upper panel shows the $V_{cb,i}=0~{\rm km}~{\rm s}^{-1}$ case, and the lower panel shows the $V_{cb,i}=70~{\rm km}~{\rm s}^{-1}$ case. At $z<20$, we slice the panels into four horizontal strips to show the cases with different $T_{\rm min}$'s ($10$, $100$, $300$, and $1000$ K) altogether.}
\label{fig:preremap}
\end{center}
\end{figure*}

With the setups described above, we create three sets of ICs with three different streaming velocity levels, $V_{bc,i}=0$, $28$, and $70~{\rm km}~{\rm s}^{-1}$. Since the streaming motion is highly coherent over a few-megaparsec scale, we implement the motion as a constant drift between gas and  dark matter within our simulation box. The probability distribution of the streaming velocity at $z=1000$ mostly ranges between 0 and $70~{\rm km}~{\rm s}^{-1}$ with it average near $\sim$ 30 ${\rm km}~{\rm s}^{-1}$
\citep[see Fig.~2 of][]{2016ApJ...830...68A}. Thus, our choices should cover most of the possible values of $V_{bc,i}$.

To generate the ICs, we use the Baryon–CDM Cosmological Initial Condition Generator for Small Scales (BCCOMICS) package of \cite{2018ApJ...869...76A}\footnote{https://github.com/KJ-Ahn/BCCOMICS}. BCCOMICS calculates the perturbation equations for the density/velocity fluctuation amplitude with the streaming velocity terms to account for the streaming effect in a self-consistent manner. The details of the calculation are described in \cite{2016ApJ...830...68A} and  \cite{2018ApJ...869...76A}. We note that the streaming motion is often implemented in an approximate manner by adding a constant drift velocity to the gas density/velocity generated without accounting for the streaming motion. In this case, the ICs do not include the smoothing effect of the streaming motion on the gas density field taking place before the initialization of the simulation, substantially overestimating the gas density fluctuation amplitude at small scales \citep{2020ApJ...900...30P}.

We show the initial density/velocity fields of the ICs in Figure~\ref{fig:ICmap}. The same set of random phases is used for both sets of ICs to prevent the cosmic variance from affecting our analysis. In the case with a streaming velocity of $70~{\rm km}~{\rm s}^{-1}$, the gas is moving toward the lower right corner of the figure with respect to the underlying dark matter. The dark matter density is nearly unaffected by the streaming, while the gas density shows some smoothing effect toward the streaming direction.

\section{Pre-reionization IGM} \label{sec:PRIGM}

We evolve the initial gas/dark matter particle distribution at $z=200$ using the publicly available smoothed particle hydrodynamics code (SPH) code GADGET-2 \citep[GAlaxies with Dark matter and Gas intEracT;][]{2001NewA....6...79S,2005MNRAS.364.1105S}\footnote{https://wwwmpa.mpa-garching.mpg.de/gadget/} down to $z=5.5$. In this step, the gas cools adiabatically as expected for the IGM before reionization.

These treatments are based upon the assumption, justified by \cite{2000ApJ...534...11H}, that the typical MH was not yet forming stars before the external I-front overtook it.  The rise of the UV ionizing radiation background responsible for the I-front that overtook each MH was led by an earlier rise of the Lyman$-$Werner background to the threshold level required to dissociate the ${\rm H}_2$ molecules inside MHs to suppress their star formation.

We first simulate structure formation with a minimum gas temperature of $T_{\rm min}=10$ K down to $z=20$. From $z=20$, we run four cases with different $T_{\rm min}$'s to simulate the impact of different levels of preheating by X-ray sources\footnote{Other effects of the X-ray background would be an increase in ionized fraction and ${\rm H}_2$ fraction in the IGM. We expect the former to be well below 10\% and not change the ionization photon budget significantly and the latter to be canceled out by the ambient Lyman$-$Werner background radiation.}. In one case, we leave $T_{\rm min}$ at $10$ K after $z=20$, assuming a minimal level of preheating. In the other three cases, we raise $T_{\rm min}$ to $100$, $300$, and $1000$ K. In combination with three levels of streaming velocity, $V_{cb,i}=0$, $28$ and $70$ ${\rm km}~{\rm s}^{-1}$, we create 12 cases of pre-reionization IGM in total. Figure~\ref{fig:preremap} describes the time evolution of the density field for $V_{cb,i}=0$ and $70~{\rm km}~{\rm s}^{-1}$.

The density maps in Figure~\ref{fig:preremap} show the lasting impacts of the streaming velocity and X-ray preheating on gas density. The case with zero streaming velocity and minimal preheating ($T_{\rm min}=10$ K) depicted in the upper side of the upper panel is the most abundant in structures, and the cases with a higher $V_{cb,i}$ or higher $T_{\rm min}$'s yields fewer structures. Notably, the streaming motion makes a substantial difference even at $z\lesssim 6$, when it has decayed below $1~{\rm km}~{\rm s}^{-1}$.  This result may appear to contradict previous studies that found that the streaming motion has a minor impact on halos that host star-forming galaxies during reionization. The mass scale for the SSGSs is below that for those galaxies ($\sim$ $10^8$ $M_\odot$), and they are still susceptible to the streaming effect. \cite{2018MNRAS.474.2173H} reported similar findings from similar simulations.

\begin{figure}
\begin{center}
\includegraphics[scale=0.5]{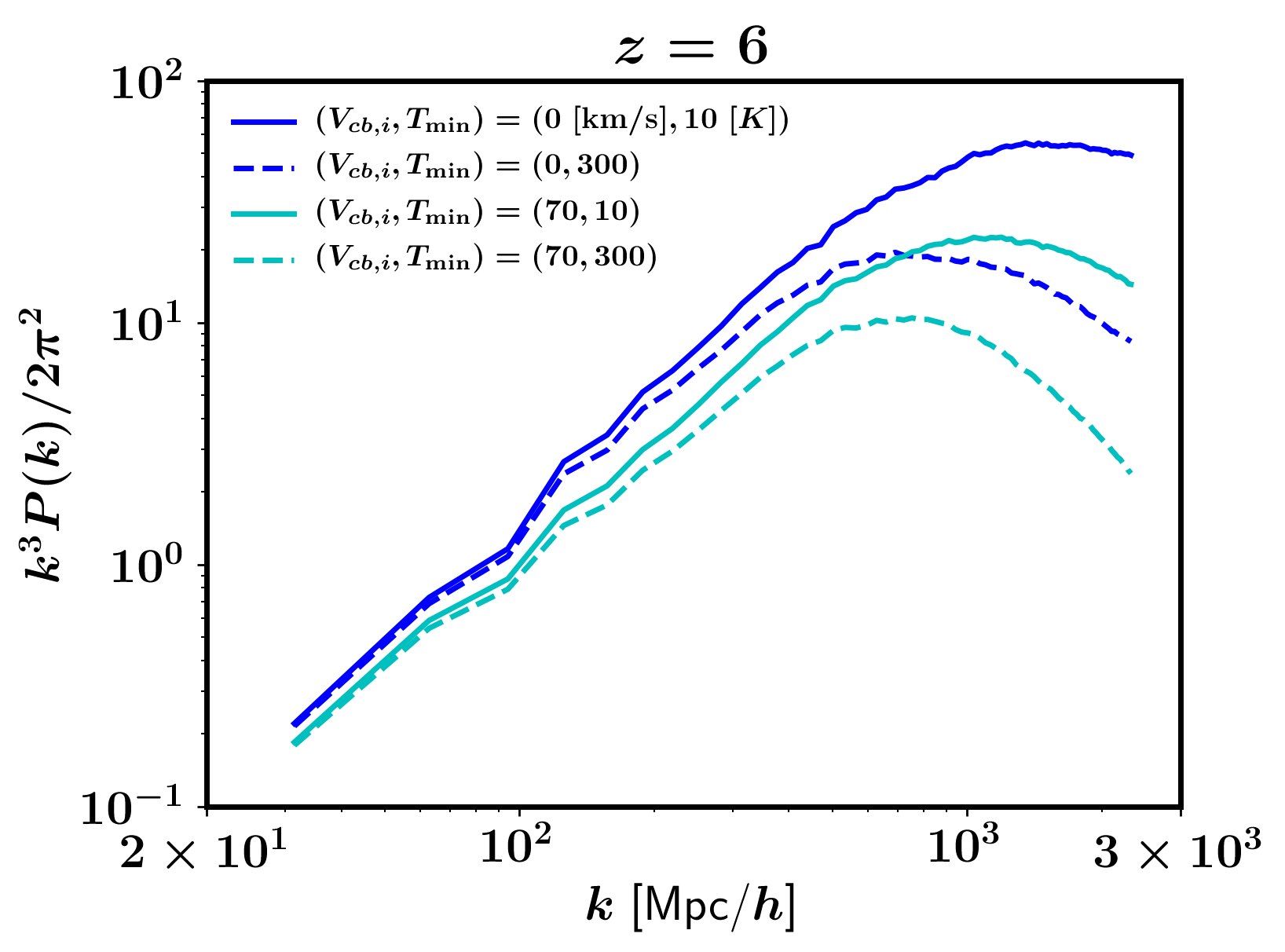}
\caption{Dimensionless gas density power spectrum of pre-reionization IGM at $z=6$. The blue and cyan lines describe the cases that $V_{cb,i}=0$, and $70~{\rm km}~{\rm s}^{-1}$, respectively. The solid, dashed, and dotted line types represent the cases where we set $T_{\rm min}=10$, $300$, and $1000~K$, respectively. }
\label{fig:PrerePS}
\end{center}
\end{figure}

We show gas density power spectra of four cases in Figure~\ref{fig:PrerePS} to describe the impacts of streaming and preheating more quantitatively. Both suppress the small-scale end of the power spectrum, but the detailed response is somewhat different between the two factors. The suppression by preheating is more focused on small scales above a particular wavenumber as the increased $T_{\rm min}$ raises the filtering mass scale, below which the structures are removed  (compare solid vs. dashed lines). On the contrary, the suppression by the streaming motion appears in a wider range of scales (compare blue vs. cyan lines).

The EOR is thought to have happened in an extended period, presumably starting from density peaks and finishing void regions in the final stage. A small volume like our simulation box will be exposed to the ionizing background at different timings, depending on where it is located in the universe. The recombination rate is expected to depend on that timing because the SSGSs would gravitationally evolve over time. Thus, we treat the reionization redshift of our simulation box ($z_{\rm re}$) as an extra parameter for the SSGS recombination rate and save the snapshots at $z_{\rm re}=5.5,~6,~7,~9$, and $12$ for each choice of $T_{\rm min}$ and $V_{cb,i}$. Therefore, we prepare 60 snapshots of pre-reionization IGM in total to simulate the impact of reionization on SSGSs.

\section{IGM after Reionization} \label{sec:IGMaR}

 \begin{figure*}
\begin{center}
\includegraphics[scale=1.1]{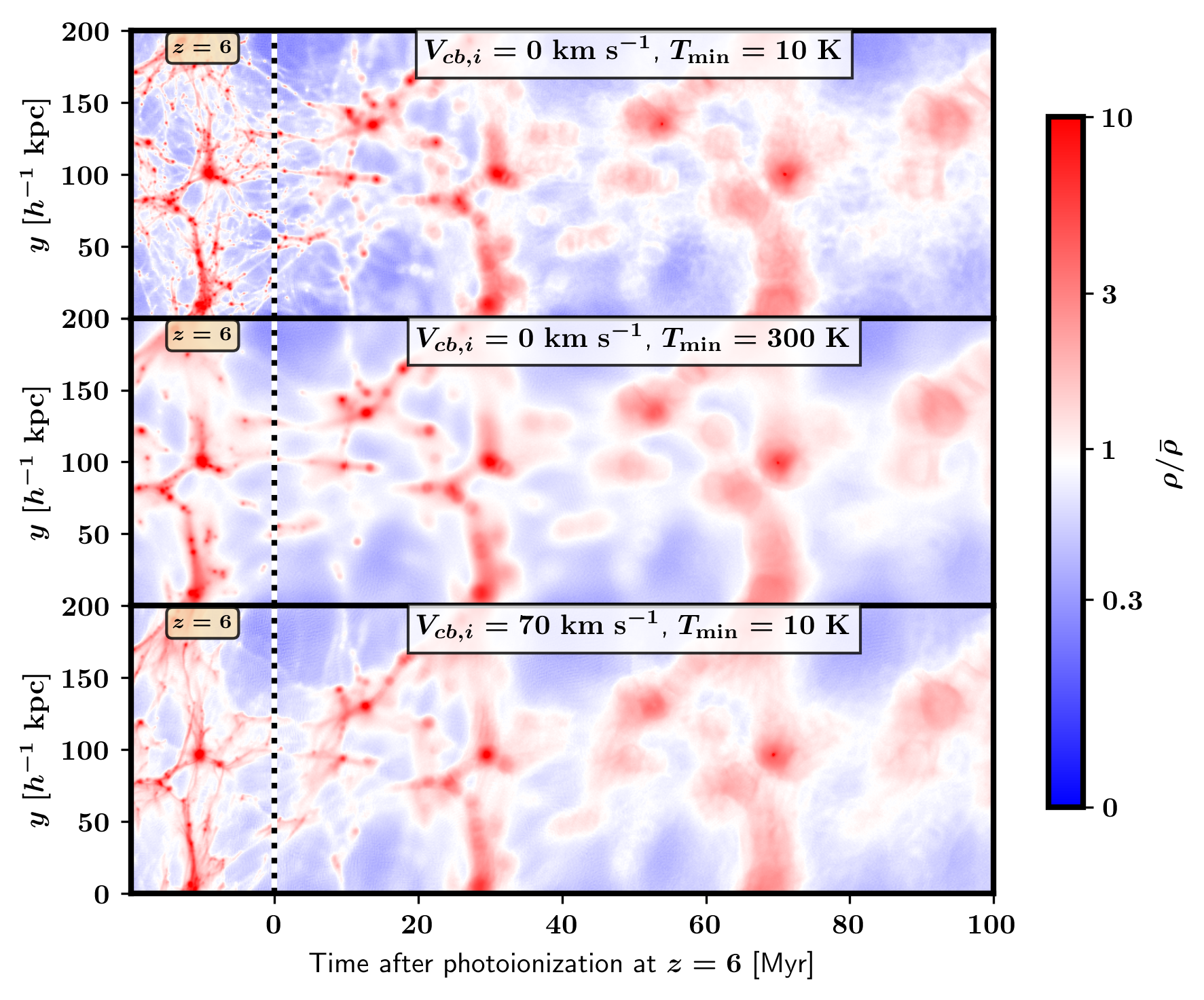}
\caption{Projected gas density maps of the three cases where EIBR is turned on at $z=6$. The three cases shown are the zero stream velocity ($V_{cb,i}=0~{\rm km}~{\rm s}^{-1}$) with minimal preheating ($T_{\rm min}=10$ K) case, zero stream velocity with a certain level of preheating ($T_{\rm min}=300$ K) case, and $2.5\sigma$ streaming velocity ($V_{cb,i}=70~{\rm km}~{\rm s}^{-1}$) with minimal preheating case in the upper, middle, and lower panels, respectively. For each panel, the left side of the vertical dotted line shows the density field at $z=6$ when we turn on the EIBR. On the right side, we stitch the density field of each snapshot to describe the time evolution of the field after the turn-on.
}
\label{fig:dmap}
\end{center}
\end{figure*}

 \begin{figure*}
\begin{center}
\includegraphics[scale=0.55]{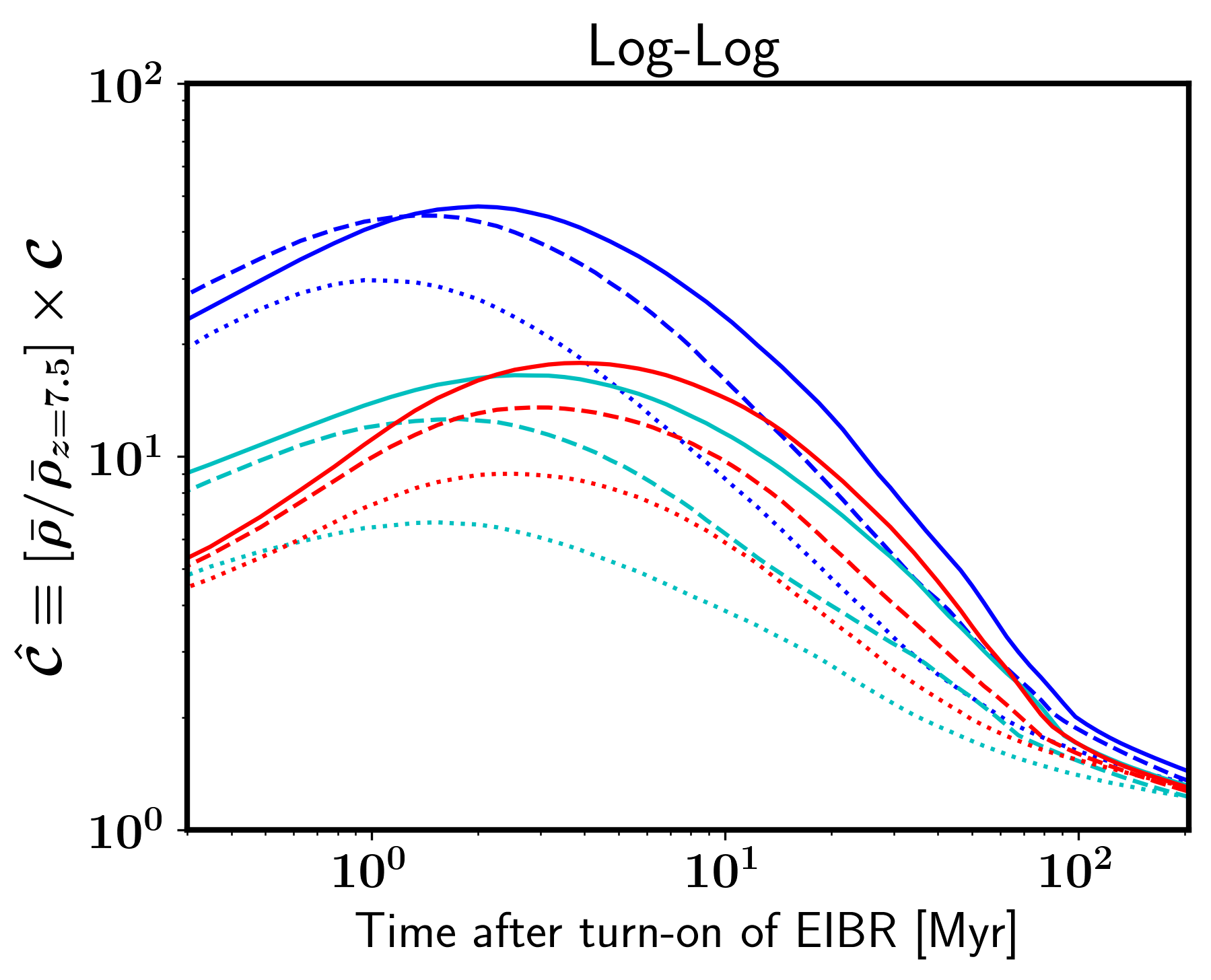}
\includegraphics[scale=0.55]{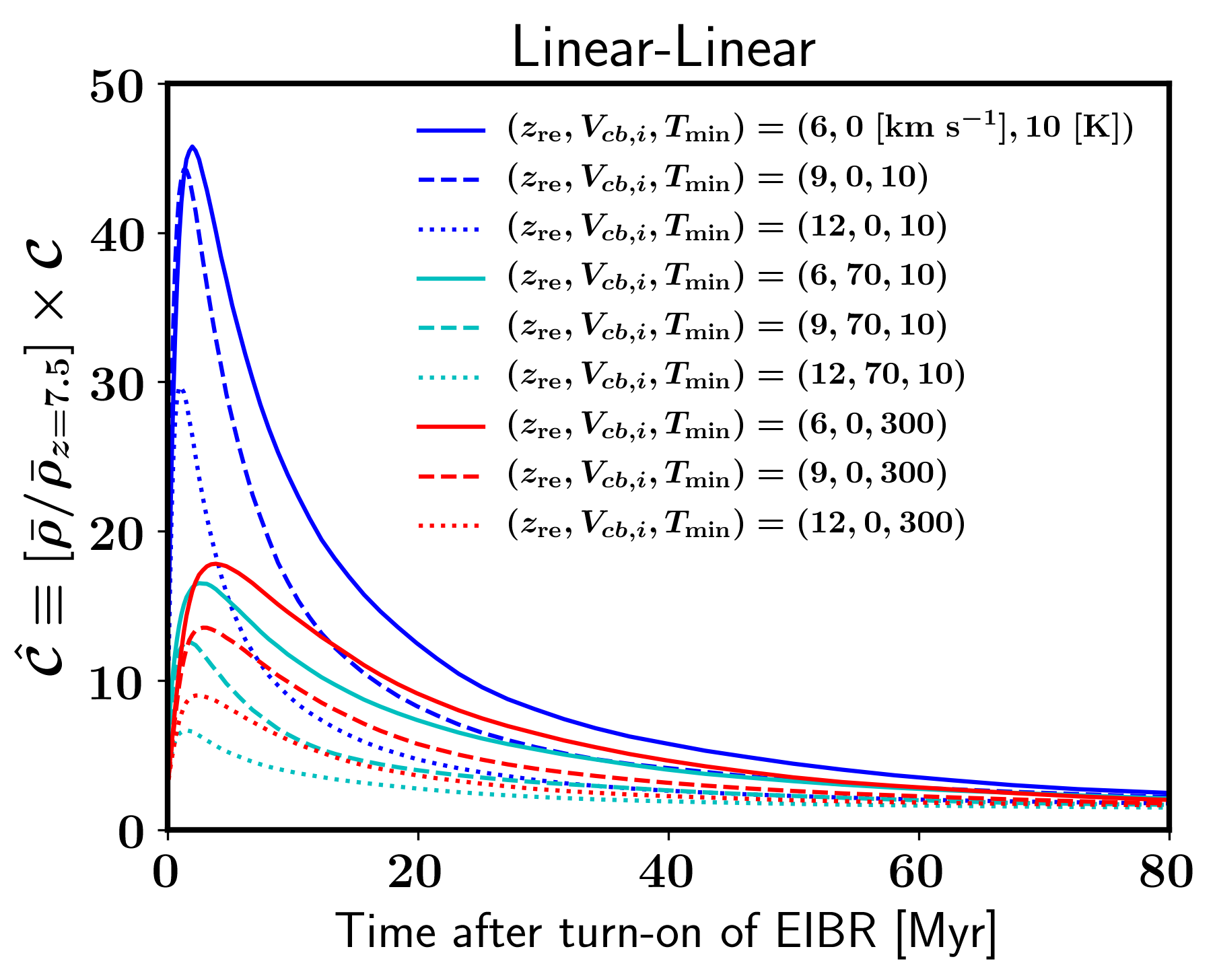}
\caption{Clumping factor multiplied by the cosmic mean density divided by its $z=7.5$ value ($\hat{\mathcal{C}}$) as the function of the time after turn-on of EIBR. The results are plotted on the log$-$log axis in the left panel and linear$-$linear axis in the right panel. The blue, cyan, and red lines describe the cases that $(V_{cb,i},T_{\rm min})=(0~{\rm km}~{\rm s}^{-1}, 10~{\rm K}),~(70~{\rm km}~{\rm s}^{-1}, 10~{\rm K})$, and $(0~{\rm km}~{\rm s}^{-1},300~{\rm K})$, respectively. The solid, dashed, and dotted lines described the cases that $z_{\rm re}=6$, 9, and 12, respectively. }
\label{fig:C}
\end{center}
\end{figure*}

We track the fate of SSGSs after the onset of the external ionizing background radiation (EIBR) in the snapshots produced from the previous steps. We use the GADGET-RT code developed from P16, where the EIBR is simulated with a self-shielding scheme and a nonequilibrium chemistry of the hydrogen and helium-related species described in \cite{2006ApJ...652....6Y,2007ApJ...663..687Y}. Among the chemical reactions included in the code, the ionization (${\rm H} \rightarrow {\rm H^{+}} + e^{-}$) and recombination (${\rm H^{+}} + e^{-} \rightarrow  {\rm H}$) of hydrogen are relevant to our calculations. We turn on the blackbody radiation of $10^5$ K with photo-ionization rate of $9.2\times 10^{-12}~{\rm cm}^{-2}~{\rm s}^{-1}~{\rm Hz}^{-1}~{\rm sr}$ or $\Gamma_{-12}=9.2$ in the pre-reionization IGM generated from the previous steps. The spectral energy distribution and radiation intensity of EIBR have been shown to have a minor impact on the evolution of SSGSs in P16, D20, and C20.

For each chemistry time step, we first list all the gas particles of which hydrogen is more than 50\% neutral. For each of those target particles, we locate the neighboring particles within 200 physical parsecs utilizing the tree structures prepared for the gravity solver. We, then, calculate the average neutral hydrogen column density for six directions, $\pm x$, $\pm y$, and $\pm z$, for each target particle. This way, we ionize neutral regions from the outskirt while shielding the inner part of the gas clump properly. P16 has shown that this radiative-transfer algorithm reproduces physical quantities of evaporating MH gas with high accuracy. The MH gas is practically the only neutral gas that is shielded against the EIBR for a significant amount of time in the simulation ($>$$10^6$ yr). 

Figure~\ref{fig:dmap} depicts the evolution of gas density after the EIBR is turned on at $z_{\rm re}=6$ for three cases. Upon reionization, the Jeans mass increases to $\sim10^8~M_\odot$, and the SSGSs begin to be destroyed by the increased pressure. Most of the structure is destroyed after a few times $10^6$ yr, while gas clumps in the MHs last up to $\sim 10^8$ yr due to self-shielding. 

P16 and D20 have shown that SSGSs substantially increase the ionization photon budget for reionization by raising the recombination rate. In the following section, we describe how we calculated the recombination rate from our simulation and how it depends on $z_{\rm re}$, $T_{\rm min}$, and $V_{cb,i}$.

\subsection{Recombination rate}

We briefly summarize how we calculate the recombination rate from our simulations and report the results. For the fully detailed derivations, we refer the readers to Section 3 of P16.

The average recombination rate in a volume $V$ is given by
\bea \label{eq:R}
\bar{\mathcal{R}}  = \mathcal{C}_{sim}\alpha_B(\bar{T}) \bar{n}_e \bar{n}_{\rm HII},
\eea
where $\alpha_{\rm B}(T)=2.6\times10^{-13}(T/10^4~K)^{-0.7}~{\rm s}^{-1}~{\rm cm}^{-3}$ is the case B recombination coefficient at temperature $T$, $\bar{n}_X\equiv\left< n_X \right>_V$ is the volume weighted averaged number density of a species $X$, and
\bea
\mathcal{C}_{sim}  =\frac{\left< \alpha_B(T)n_e n_{\rm HII}\right>_V}{\alpha_{\rm B}(\bar{T}) \bar{n}_e \bar{n}_{\rm HII} },
\eea
is the clumping factor, which accounts for the boost in the recombination rate due to the spatial fluctuations in $T$, $n_e$, and $n_{\rm HII}$. Here, the subscript $S$ denotes that the results from our simulations only include clumping in SSGSs.
The mean temperature $\bar{T}$ is given by
\bea
\bar{T}=(m_p/k_B)\left< (\gamma -1 )u \right>_M \left< \mu^{-1} \right>^{-1}_M,
\eea
where $u$ is the specific internal energy, $\mu$ is the mean molecular weight, $\left<\right>_M$ denotes the mass-weighted average. We note that, as was reported in P16, our definition of the clumping factor gives about a 10\% higher recombination rate than the conventional one, $\left< n^2 \right> /\left<n\right>^2$, does because a higher cooling rate in a higher density makes $\alpha_B(T)$ mildly correlate with density. 

Nearly 100\% of the hydrogen atoms are ionized shortly after the onset of EIBR except for the dense MH gas, which takes only a small fraction of total gas giving $n_{\rm HII} \approx n_H$ and helium atoms are singly ionized giving 8\% of extra free electrons: $n_e \approx  (1.08)n_H$. Thus, the evolution of the clumping factor is the major factor that dominates the recombination rate. In the case of the SPH particle data, the clumping factor can be computed from the following equation:
\bea
\mathcal{C}_{sim} =\bar{n}^{-1}N_{\rm ptl} \left[  \frac{\Sigma_i f_{e,i} f_{{\rm H II},i} \alpha_{\rm B}(T_i) n_i}{ \left( \Sigma_i f_{e,i} \right)  \left( \Sigma_i f_{{\rm H II},i} \right) \alpha_{\rm B}(\bar{T}) } \right]
\eea
Here, the subscript $i$ denotes the $i$th SPH particle in the simulation, $N_{\rm ptl}$ is the number of SPH particles, $n\equiv \rho/m_p$ is density in the unit of the proton mass, $f_X\equiv n_X/n$ is the number density of a species $X$ divided by $n$, and $\bar{T}$ is given by averaging over the particle values: $N_{\rm ptl}^{-1}\Sigma_iT_i$.

Due to cosmic expansion, volumes with the same clumping factor yield different recombination rates at different redshifts. To compare the recombination rate from different $z_{\rm re}$`s, it is convenient to weight the clumping factor by the cosmic mean density $\bar\rho$. Specifically, multiplying the clumping factor by $\bar\rho$ normalized at its $z=7.5$ value ($\hat{\mathcal{C}}_{sim} \equiv[\bar\rho/\bar\rho_{z=7.5}]\mathcal{C}_{sim} $) makes $\hat{\mathcal{C}}_{sim} =1$ correspond to the recombination rate of 1 per H atom per gigayear. We plot $\hat{\mathcal{C}}_{sim} $ for nine cases in Figure~\ref{fig:C}.

On a log$-$log axis (left panel of Fig.~\ref{fig:C}), the clumping factor as the function of time shows a rise and decay as reported in P16 and D20. Initially, the ionization fronts are in the supersonic $R$-type phase, where the ionization proceeds toward dense gas clumps before the gas responds to the photoheating. The clumping factor rises as denser gas is ionized. After several megayears, ionized gas begins to expand and lower the clumping factor. On the linear-linear axis (right panel of Fig.~\ref{fig:C}), the early $R$-type phase is negligibly short and unimportant for the integrated recombination, which is proportional to the area under the curve. 

Comparison of $\hat{\mathcal{C}}_{sim}$ in the nine cases shown in Figure~\ref{fig:C} confirm that the clumping factor depends sensitively on the three parameters, $z_{\rm re}$, $V_{cb,i}$, and $T_{\rm min}$: $\hat{\mathcal{C}}_{sim}$ is higher for lower $z_{\rm re}$'s, lower $V_{cb,i}$'s, and lower $T_{\rm min}$'s. These parameter dependences have been partially shown in previous studies. D20 ran one case with $T_{\rm min}=1000$ K to find a much lower clumping factor than in the $T_{\rm min}=10$ K cases, C20 showed that the clumping factor is suppressed when the streaming velocity is higher, and these studies repeatedly showed the $z_{\rm re}$  dependence. This study focuses on computing the variation in the reionization history due to the spatial fluctuations in $V_{cb,i}$ and $T_{\rm min}$.

\section{Reionization history} \label{sec:RHist}

\begin{table*}
\centering
\caption{Reionization Source Models and variation in the End-of-reionization redshift}
\label{tab:sim}
\begin{tabular}{ccccccc}
Model        & $f_S$  & $f_L$ &  $z_{e}$ ($T_{\rm min}=10~K$) &  $z_{e}$ ($T_{\rm min}=100~K$) & $z_{e}$ ($T_{\rm min}=300~K$) & $z_{e}$ ($T_{\rm min}=1000~K$)   \\ \hline
Accelerating & 0 & $1560$     &       $5.76-6$\footnote{The range is given by bracketing the results from the no-streaming case and maximal-streaming case.} & $5.84-6.04$ & $5.94-6.08$ & $6.09-6.15$ \\ \hline
Decelerating & 0 & $1560[z/7.21]^4$ & $5.27-6$ & $5.53-6.10$ & $5.83-6.20$ & $6.22-6.33$ \\ \hline
Self-regulated & 15600  & $975$  &    $5.78-6$ & $5.86-6.04$ & $5.94-6.08$ & $6.09-6.15$ \\ \hline
\end{tabular}\label{tab:sources}
\end{table*}

Here, we calculate the reionization history from a one-zone model similar to the one used by D20 to assess the impacts of the streaming motion and the X-ray preheating. Due to the long fluctuation scale of the streaming motion ($\sim 150$ Mpc), we expect volumes that are $\lesssim 50$ Mpc across to form a zone where $V_{cb,i}$ is more or less uniform. We intend to describe the difference in reionization histories in such zones with different local values of $V_{cb,i}$.

We emphasize that our results from the one-zone model should {\it not} be taken to be quantitatively accurate as it leaves out most of the complexities of reionization. For example, the spatial inhomogeneity of reionization due to clustering of ionizing sources cannot be modeled by design. We defer more accurate modelings of reionization to future studies with more sophisticated numerical simulations.

\subsection{One-zone Model} 

The change in the global mean ionization fraction of the IGM ($Q_{\rm HII}$) is given by the volume-averaged difference between the ionization rate $\mathcal{I}$ and the recombination rate $\mathcal{R}$:
\bea\label{eq:Q}
\frac{dQ_{\rm HII}}{dt}=
\bar{n}_{\rm H}^{-1}(\bar{\mathcal{I}} - \bar{\mathcal{R}})
\eea
\citep{1987ApJ...321L.107S,1999ApJ...514..648M}. The volume average here involves the entire universe, unlike in the previous section where we only averaged over the simulation box. In this case, Equation~(\ref{eq:R}) is slightly changed to
\bea
\bar{\mathcal{R}} &=& \mathcal{C}_{tot} \alpha_B(\bar{T}) \bar{n}_e \bar{n}_{\rm HII} \nonumber\\
&=& \mathcal{C}_{tot} \alpha_B(\bar{T})(1.08) \bar{n}_{\rm H}^2 Q_{\rm HII} ,
\eea
where 
\bea
\mathcal{C}_{tot}&\equiv&[\mathcal{C}_{L}-1]+[\mathcal{C}_{S}-1]+1 \nonumber \\
&=& \mathcal{C}_{L}+\mathcal{C}_{S}-1
\eea
combines clumping from large scales beyond our simulation box ($\mathcal{C}_{L}$) and that from SSGSs in the simulation ($\mathcal{C}_{S}$). The large-scale contribution comes from structures above the Jeans mass of ionized gas and can be obtained from large-scale simulations with the box size $\gtrsim$ 10 Mpc. In this study, we adopt the fitting function from \citet{2009MNRAS.394.1812P}: $\mathcal{C}_{L}=1+43z^{-1.71}$. For the global SSGS contribution $\mathcal{C}_{S}$, one must consider that different regions in the universe can be exposed to the EIBR at different $z_{\rm re}$'s. Thus, we average the simulated results of $\mathcal{C}_{sim}$ over the past reionization history:
\bea
&&\mathcal{C}_{S}(z)= \nonumber \\
&&\frac{\int^z_{z_i} \dot{Q}_{\rm HII}(z^\prime)\mathcal{C}_{sim}(\Delta t^\prime, z^\prime)(d\Delta t^\prime/dz^\prime)dz^\prime }{Q_{\rm HII}(z)},
\eea
where $\Delta t^\prime=t(z)-t(z^\prime)$ is the time difference between $z$ and $z^\prime$. We assume that the reionization begins at $z=12$ and interpolate the simulated results of $\mathcal{C}_{sim}(\Delta t, z_{\rm re})$ obtained for $z_{\rm re}=5.5,~6,~7,~9,$ and $12$ to evaluate the integral. 

 \begin{figure*}
\begin{center}
\includegraphics[scale=0.41]{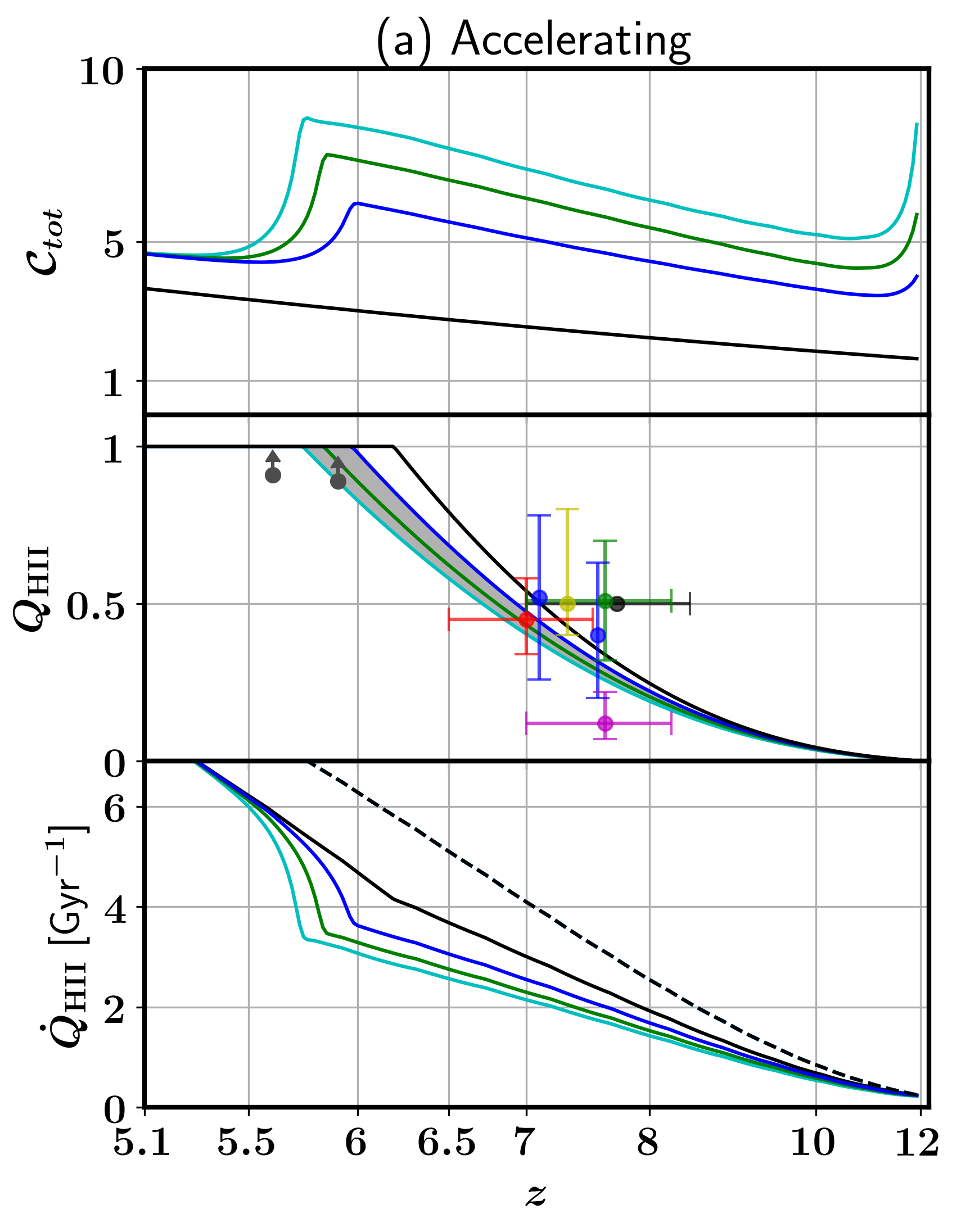}
\includegraphics[scale=0.41]{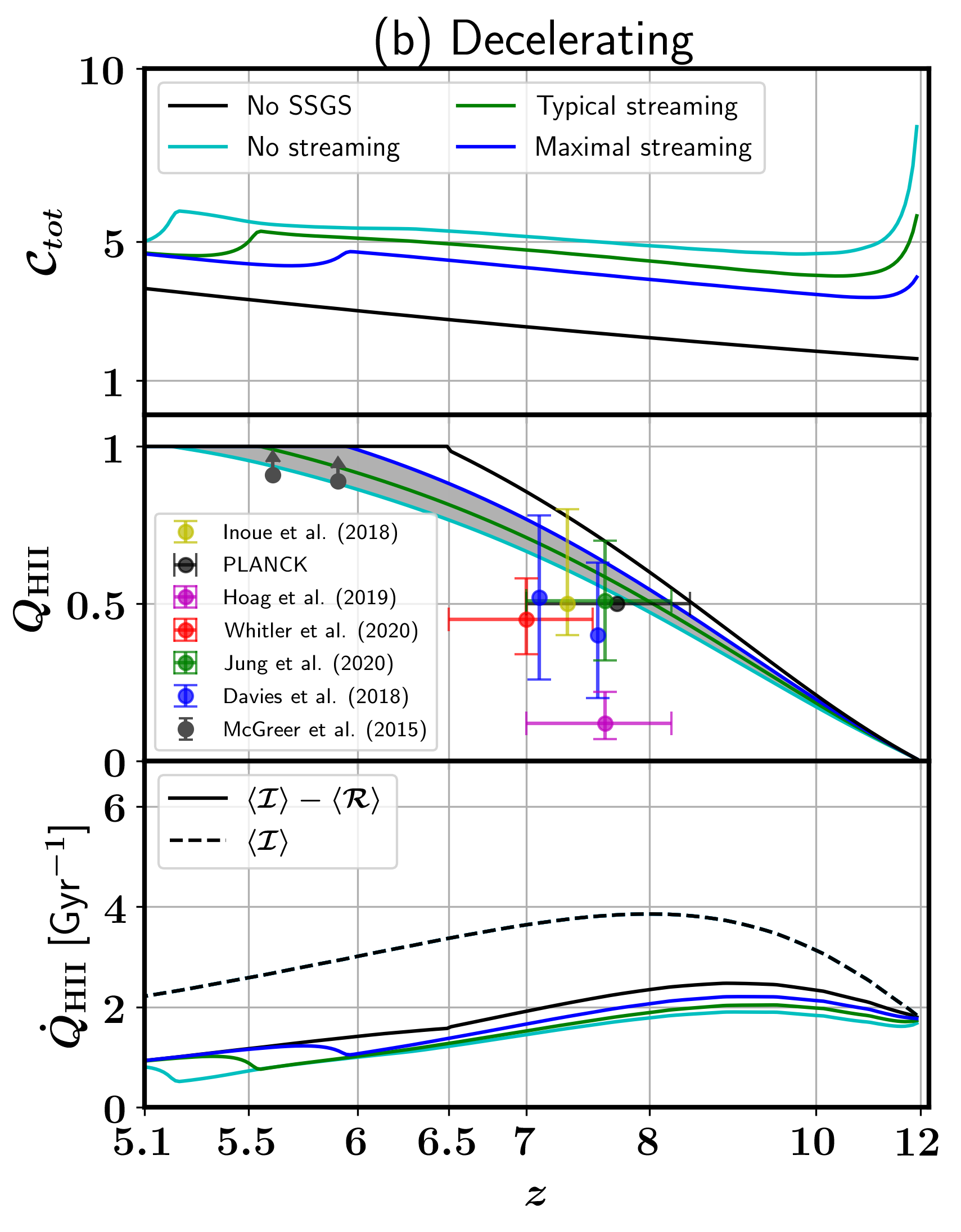}
\includegraphics[scale=0.41]{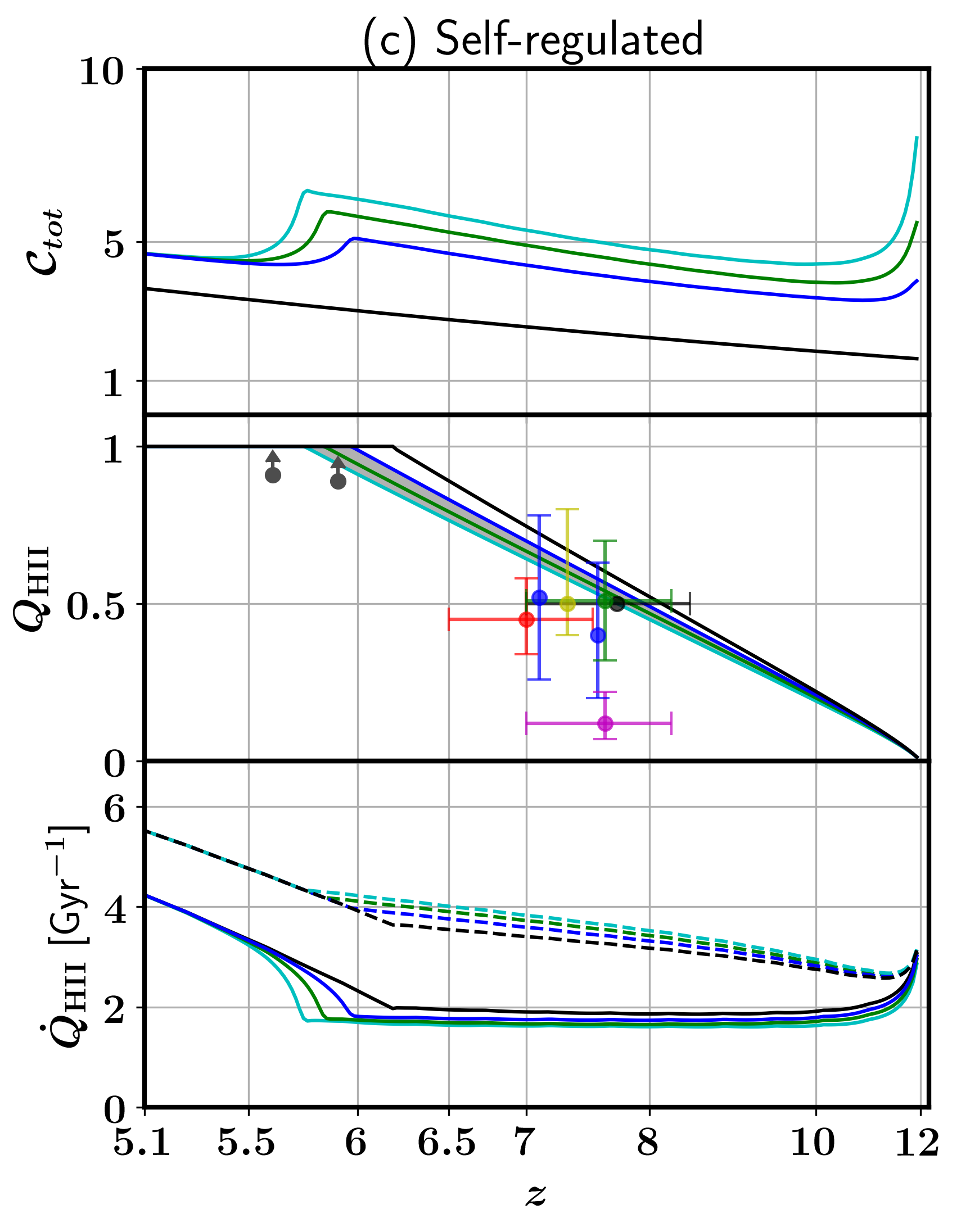}
\caption{Average clumping factor (upper), ionization fraction (middle), and ionization rate (lower) as functions of time. The black lines describe the no-SSGS case. The cyan, green, and blue lines describe the cases where the recombination rate is computed for the no-streaming, typical-streaming, and maximal-streaming zones, respectively. The dashed line represents the production rate of the ionizing photons. In the middle panels, the observational constraints on the reionization history from \citet{2018ApJ...864..142D}, \citet{2018PASJ...70...55I}, \citet{2018arXiv180706209P}, \citet{2019ApJ...878...12H}, \citet{2020MNRAS.495.3602W}, and \citet{2020ApJ...904..144J} are shown as blue, yellow, black, magenta, red, and green data points, respectively. The lower bounds from \citet{2015MNRAS.447..499M} are shown in gray. Panel $(a)$, $(b)$, and $(c)$ describe the accelerating, decelerating, and self-regulated reionization models, respectively. }
\label{fig:RHist}
\end{center}
\end{figure*}

For the source term of Equation~(\ref{eq:Q}), $\bar{n}_{\rm H}^{-1}\bar{\mathcal{I}} $, we construct our one-zone models motivated by \cite{2007MNRAS.376..534I} using two classes of ionizing sources, the low-mass sources ($\lesssim 10^9~M_\odot$) whose formation is suppressed in H II regions due to Jeans mass filtering and the high-mass sources ($\gtrsim 10^9~M_\odot$) that are not suppressed. Then, the ionization term is given by
\bea \label{eq:I}
\bar{n}_{\rm H}^{-1}\bar{\mathcal{I}}=f_H+(1-Q_{\rm H II}^n)f_L,
\eea
where $f_L$ and $f_H$ are the efficiencies of the low- and high-mass sources, respectively. Here, the factor, $1-Q_{\rm H II}^n$, accounts for the suppression of low-mass sources in ionized regions. The exponent $n$ would equal 1 if the low-mass sources were homogeneously distributed in space, but, in practice, it is less than $1$ because the low-mass sources are highly clustered around density peaks where the reionization begins, and galaxies are more efficiently suppressed at the early time than at the late time of the reionization. \cite{2007MNRAS.376..534I} empirically found that $n=0.1$ fits their simulated results. They provide the fitting functions for the $f_L$ and $f_H$ as the following\footnote{Equations~(A2) and (A3) of their work.}:
 \bea
f_L&=&0.335\left( \frac{f_{\gamma,L}}{250} \right) \exp(0.227z-0.02546z^2)~{\rm Myr}^{-1} \\
f_H&=&4.6312\left( \frac{f_{\gamma,H}}{250} \right) \exp(-0.107z-0.02463z^2)~{\rm Myr}^{-1}, \nonumber\\
\eea
Here,  the $z$ dependence is from the evolving source formation rate over the cosmic history, and $f_{\gamma,L}$ and $f_{\gamma,H}$ are the efficiency parameters for the two classes of ionizing sources.

We consider three models for the source term $\bar{n}_{\rm H}^{-1}\bar{\mathcal{I}}$ by modulating $f_{\gamma,L}$ and $f_{\gamma,H}$ as in Table~\ref{tab:sources}. Descriptions of the models are as follows.
\begin{itemize}
\item[1.] Accelerating model: Only consider the contribution from high-mass sources ($f_{\gamma,E}$) , which are not suppressed by ionizing radiation. The source term monotonically increases over time, thereby accelerating the reionization and is qualitatively similar to the \citet{2015ApJ...802L..19R} model. 
\item[2.] Decelerating model: Multiply a factor of $(z/7.21)^4$ to the accelerating model to make the sources less efficient at a later time, thereby decelerating the reionization. This is motivated by the ``early reionization" scenario of \citet{2019ApJ...879...36F}, where the escape fraction of ionizing photons decreases over time.
\item[3.] Self-regulated model: Includes the contribution from low-mass sources ($f_{\gamma,L}$), which are suppressed as the global ionization fraction $Q_{\rm HII}$ rises. This scenario assumes that the low-mass sources have a much higher ionizing radiation escape fraction than the high-mass ones do, allowing them to contribute significantly to the ionizing background \citep[see, e.g.,][]{2020arXiv200813215H}.
\end{itemize}
For each model, the values of $f_{\gamma,L}$ and $f_{\gamma,H}$ are tuned so that the reionization finishes at $z=6$ when $V_{cb,i}=70~{\rm km}~{\rm s}^{-1}$ and $T_{\rm min}=10$ K. 

For the sink term $\bar{n}_{\rm H}^{-1}\bar{\mathcal{R}}$, we consider three ``zones" with different recombination rates: 
\begin{itemize}
\item[1.] No-streaming zone: The small-scale clumping factor $\mathcal{C}_S$ is obtained from simulations with $V_{cb,i}=0~{\rm km}~{\rm s}^{-1}$ to represent volumes with a minimal streaming motion. 
\item[2.] Typical-streaming zone: To represent the volumes with typical magnitude of streaming velocity, $\mathcal{C}_S$ is obtained from simulations with $V_{cb,i}=28~{\rm km}~{\rm s}^{-1}$.
\item[3.] Maximal-streaming zone: $\mathcal{C}_S$ is obtained from simulations with $V_{cb,i}=70~{\rm km}~{\rm s}^{-1}$, which give a maximal impact by the streaming motion. 
\item[4.] No-SSGS zone: A hypothetical case where the SSGS contribution is ignored by setting $\mathcal{C}_S=0$.
\end{itemize}


\subsection{Variation in reionization history induced by the streaming motion} \label{sec:ScatterRHist}

\begin{figure*}
\begin{center}
\includegraphics[scale=0.6]{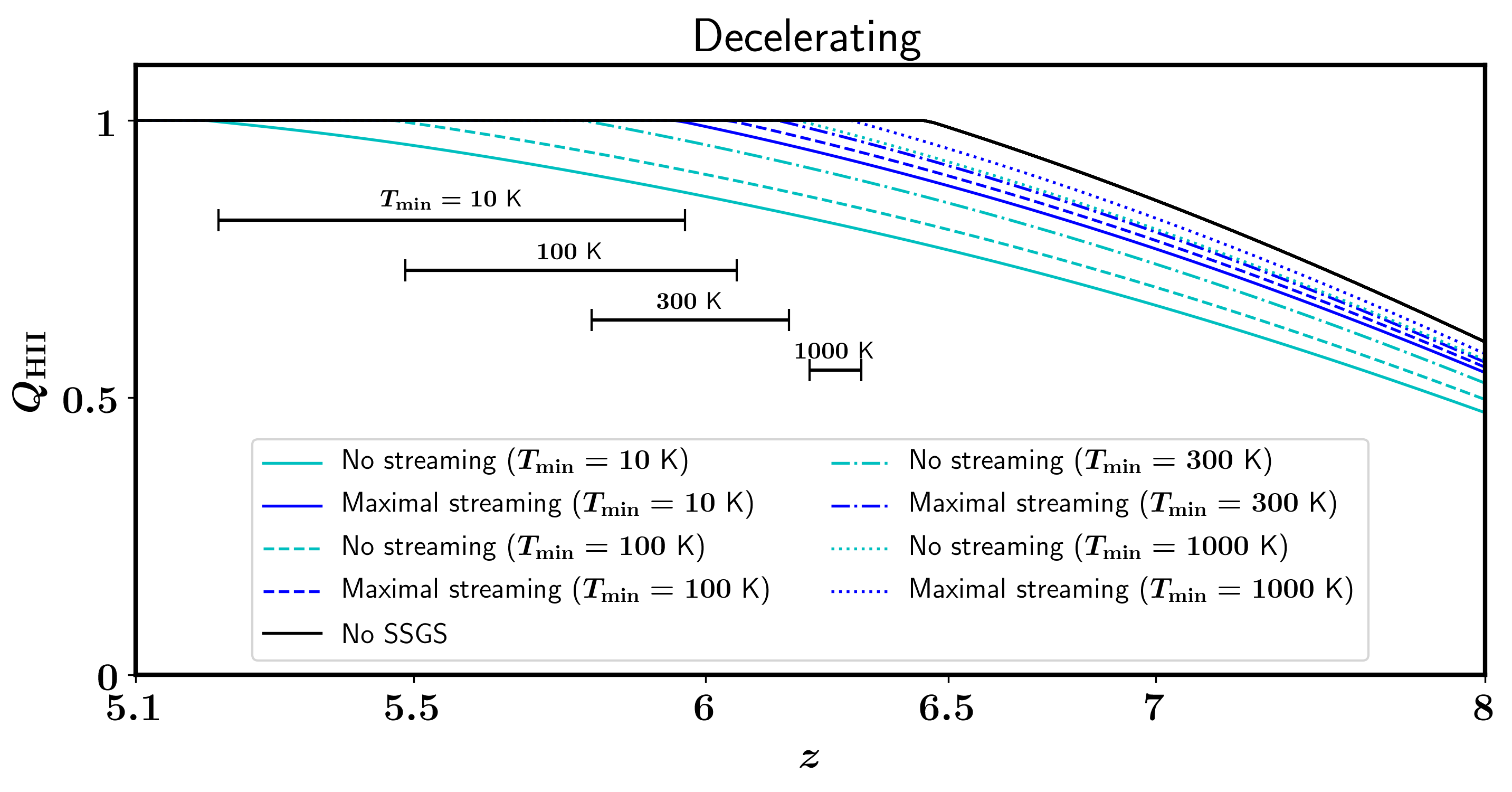}
\caption{Reionization history $Q_{\rm HII}(z)$ of no-streaming (cyan) and maximal-streaming (blue) zones for different levels of X-ray preheating for the decelerating reionization model. Solid, dashed, dotted$-$dashed, and dotted lines describe the $T_{\rm min}=10$, $100$, $300$, and $1000$ K cases, respectively. The no-SSGS case is shown as the black solid line. The black horizontal line-segments describes the $z_e$ range bracketed by the no-streaming and maximal-streaming cases for each choice of $T_{\rm min}$.}
\label{fig:Q_vs_T}
\end{center}
\end{figure*}

We first consider the minimal X-ray preheating cases with $T_{\rm min}=10~{\rm K}$.  We present the clumping factor $\mathcal{C}_{tot}$ (upper panels), the reionization history $Q_{\rm HII}(z)$ (middle panels), and the net ionization rate $dQ_{\rm HII}/dt$ (lower panels) in Figure~\ref{fig:RHist} for the three cases of the sink term with three reionization sources models. We assume that the results from the no-streaming zone and the maximal-streaming zone bracket the possible scatter in reionization history caused by the streaming motion and show that range as the gray shading. The variation in the end-of-reionization redshift $z_e$ for $T_{\rm min}=10~{\rm K}$ is listed in Table~\ref{tab:sources}.

The no-SSGS case generally finishes reionization much earlier than the no-streaming zone does. In the decelerating model (Fig.~\ref{fig:RHist}b), for example, reionization ends at $z_e\approx5.3$ in the no-streaming zone, while it ends at $z_e\approx6.5$ in the no-SSGS zone. This highlights the significant contribution from SSGSs to the ionization photon budget, which has been reported by P16 and D20. 

The reionization history of the maximal-streaming zone falls near the midpoint between those of the other two cases ($z_e=6$), showing that the streaming motion can undo a significant portion of the delay in reionization caused by SSGSs. The main interest of this work is the difference between the no-streaming zone and the maximal-streaming zone, which sets the maximum impact of the streaming motion on the reionization history. The reionization history of the typical-streaming zone falls near the midpoint between those of the no-streaming and the maximal-streaming zones, suggesting that the impact of the streaming velocity is more or less linearly proportional to its initial magnitude $V_{cb,i}$.

Since the maximal-streaming zone is tuned to give $z_e=6$, $z_e$ in the no-streaming zone sets the scatter in reionization history, which are 5.76, 5.26, and 5.78 in the accelerating, decelerating, and self-regulated reionization models. These are comparable to the $z_e=5.5$ inferred from ULAS J0148+0600, suggesting that the streaming motion can help explaining the apparent variation in Ly$\alpha$ opacity.

The clumping factor $\mathcal{C}_{tot}$ in the upper panel of Figure~\ref{fig:RHist} shows what fraction the recombination is suppressed by the streaming motion. The clumping factor contribution from SSGSs in the no-streaming (maximal-streaming) zone is given by the gap between cyanv(blue) and black lines. The SSGS contribution in the maximal-streaming zone is roughly half of that in the no-streaming zone in all three source models, showing that the streaming motion cancels about half of the recombination by SSGSs.

\subsubsection{Reionization Source Model Dependence}

The characteristic of each source model is well described by the ionization source term $\left< \mathcal{I} \right>$ (lower panels of Fig.~\ref{fig:RHist}). The source term in the accelerating model starts at $\sim 0.1~{\rm Gyr}^{-1} $ and increases steeply with time, while it decreases after peaking at $2~{\rm Gyr}^{-1} $ at $z=8$ in the decelerating model. In the self-regulated model, the ionizing efficiency rises slowly, falling between the other two models. 

A notable feature of the self-regulated model is that the source term is larger when the sink term is larger, partially canceling out the difference in net ionization rate caused by different amounts of recombination. This is because the ``self-regulation" mechanism suppresses more ionizing sources when the global ionization fraction is higher.

The characteristic of each source model explains the resulting reionization history (see middle panels of Fig.\ref{fig:RHist}). The global ionization fraction in the accelerating model rises later than in other models, reaching 50\% at $z\approx 7$ while it does at $z\approx 8$ in the other two models. Due to the steep rise of the ionizing efficiency, the global ionization fraction in the accelerating model catches up with that in the other two models near the end of reionization. The reionization history of the accelerating models agrees well with the constraint from \citet{2018ApJ...864..142D} and \citet{2020MNRAS.495.3602W}, while the other two models agree better with the constraints from \citet{2018PASJ...70...55I}, \citet{2018arXiv180706209P}, \citet{2020MNRAS.495.3602W}, and \citet{2020ApJ...904..144J}. The constraint on $Q_{\rm HII}$ from \citet{2019ApJ...878...12H} is lower than in all the three models considered in this work. The lower bounds on $Q_{\rm HII}$ at $z<6$ from \citet{2015MNRAS.447..499M} are consistent with all the cases considered in this work.

The variation in reionization history caused by the streaming motion is notably larger in the decelerating model than in other models: $z_e$ ranges between 5.27 and 6 in the decelerating model, while it ranges between $\sim5.75$ and 6 in the other two models (see Table~\ref{tab:sources}). This shows how sensitively the variation depends on the characteristic of reionization sources. Below, we explain how the source model affects the variation.

We find that a hydrogen ion in the no-streaming zone generally recombines, on average, about 0.15 times more than in the maximal-streaming zone by $z=6$. As a result, the global ionization fraction in the no-streaming zone is near 85\% in the accelerating model and the decelerating model, while 92\% in the self-regulated model, where the self-regulation partially cancels the difference in ionization rate. The steeply rising source term in the accelerating model finishes the remaining neutral gas in the no-streaming zone relatively quickly. The reionization in the self-regulated model is finished as early as in the accelerating model despite the lower ionizing rate because of the lower remaining neutral fraction. On the contrary, the end of reionization in the no-streaming zone is much delayed by the declining ionizing rate after $z=6$ in the decelerating model. After $z=6$, a few percent of IGM neutral fraction persists to $z\lesssim 5.3$ in the decelerating model, being consistent with the constraints in \cite{2015MNRAS.447..499M}.

Thus, we conclude that the ionizing photon production rate after $z=6$ is a critical factor in the variation in the reionization history caused by the streaming motion as well as the difference in local streaming velocity magnitude. The variation would increase further if we lowered the ionizing efficiency at $z<6$ in our source model.

\subsection{Impact of X-Ray preheating} 

To assess the impact of X-ray preheating, we calculate reionization history for four different minimum gas temperatures of $T_{\rm min}=10,~100,~300,$ and $1000~{\rm K}$. The results are shown in Figure~\ref{fig:Q_vs_T} for the decelerating source model. The $z_e$ variations for these $T_{\rm min}$'s are listed in Table~\ref{tab:sources} for all the three reionization source models considered in this work.

As $T_{\rm min}$ increases, the impact of the streaming motion diminishes: the reionization histories in both the maximal-streaming and no-streaming zones converge toward the no-SSGS case, giving $\Delta z_e=0.48$, 0.38, 0.26, and 0.07 for $T_{\rm min}=10,~100,~300,$ and $1000~{\rm K}$, respectively. In the most extreme X-ray preheating case of $T_{\rm min}=1000~{\rm K}$, the recombination from SSGSs is nearly wiped out.

The effect of the X-ray preheating is similar to that of the streaming motion: suppressing the SSGSs (see Fig.~\ref{fig:preremap}). Therefore, raising the preheating level globally would only the reduce the impact of the streaming motion.

\begin{figure*}
\begin{center}
\includegraphics[scale=0.5]{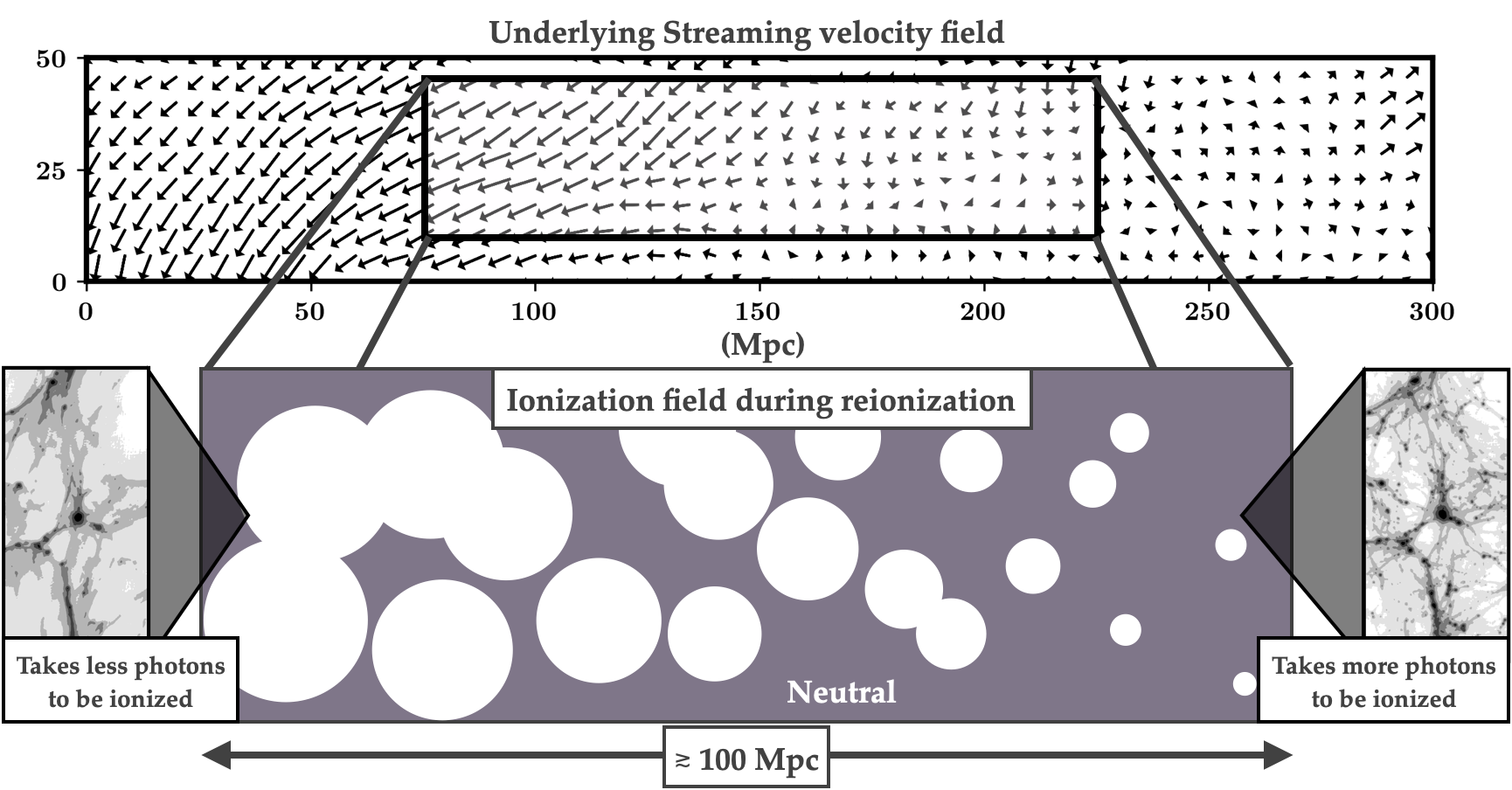}
\caption{Schematic description of how the streaming velocity (upper panel) can generate a large-scale inhomogeneity in the ionization field (bigger bottom panel). The streaming velocity field in the upper panel is an actual realization by BCCOMICS. The two small panels in the bottom are the simulated pre-reionization SSGSs at $z=6$ with $(T_{\rm min},V_{cb,i})=(10~{\rm K},70~{\rm km}~{\rm s}^{-1})$ (left) and $(10~{\rm K},0~{\rm km}~{\rm s}^{-1})$ (right).}
\label{fig:concept}
\end{center}
\end{figure*}

\section{Summary and Discussion} \label{sec:discussion}

Since \citet{2010PhRvD..82h3520T} pointed out the potentially large impact of the relative streaming motion between baryons and dark matter on early structure formation, a number of studies followed to assess its impact on early galaxy formation and reionization. In this study, we have investigated the possibility of the streaming motion affecting the reionization {\it sink}, the small-scale gas clumping in the pre-reionization IGM. We have shown that the streaming motion can erase a significant portion of the clumpiness in the pre-reionization IGM and thus make reionization happen significantly earlier under certain conditions. Because the streaming velocity varies over large length scales ($\gtrsim 100$ Mpc), our results suggest that the streaming motion can contribute to the variation in Ly$\alpha$ opacity observed from different quasar sightlines, and help to explain the large scatter from recent observation when combined with fluctuations in the UV background and IGM temperature. A schematic description of the large-scale inhomogeneity caused by the streaming motion is given in Figure~\ref{fig:concept}.

A hydrogen ion in a no-streaming ($V_{cb,i}=0~{\rm km}~{\rm s}^{-1}$) region can recombine $\sim 0.15$ times more than in a highly streaming ($V_{cb,i}=70~{\rm km}~{\rm s}^{-1}$) region throughout the reionization era. As a result, 15\% of the IGM remains neutral in the no-streaming zone when the reionization is already finished in the streaming zone. In our one-zone model for reionization history, we find that the end-of-reionization redshift can vary by up to $\Delta z_e\sim 0.5$ depending on the characteristic of ionizing sources.  We find that the variation increases as the ionizing efficiency of sources declines toward the end of reionization. On the other hand, the variation is much smaller when the source efficiency does not evolve, and the ionizing background steeply rises at late time. The variation can also be suppressed if low-mass galaxies are self-regulated due to the photoionization feedback.

X-ray preheating is another critical factor that affects the reionization history. The current lack of knowledge about the pre-reionization IGM temperature adds a large uncertainty to the recombination by SSGSs. Raising the minimum gas temperature $T_{\rm min}$ results in the suppression of SSGSs, and therefore, less recombination by SSGSs. In the most extreme case of $T_{\rm min}=1000~{\rm K}$, the preheating almost completely suppresses the recombination by SSGSs. However, we note that this conclusion is based on the assumption that the X-ray preheating is spatially homogeneous. The X-ray background may have had a significant level of spatial fluctuations generated by the large-scale distribution of X-ray sources. In that case, the fluctuating X-ray background can add to the variation in $z_e$. An interesting possibility is that the streaming velocity induces intermediate-mass direct collapse black holes as suggested by \cite{2017Sci...357.1375H}, and the radiation from these black holes constitutes a significant portion of the  X-ray background. In such cases, the X-ray background would correlate with the streaming velocity magnitude, possibly enhancing the variation in reionization history. Hence, the X-ray preheating is another critical, yet uncertain factor that determines the reionization history. 

We emphasize that this study aims to highlight the potential importance of the streaming velocity in a qualitative sense. The numerical values from our one-zone models with our parameter choices are subject to numerous uncertainties. The validity of our assumption that the reionization begins at $z=12$ is largely subject to the MHs contribution in reionization, which may ionize up to 15\% of the IGM by that redshift \citep[see, e.g.,][]{2012ApJ...756L..16A}. Also, the two levels of streaming velocity that bracket the entire range, $V_{cb,i}=0$ and $70~{\rm km}~{\rm s}^{-1}$, are unlikely extremes for the mean value of volumes that are $\sim 50$ Mpc across. A more realistic comparison would involve values like $V_{cb,i}\sim 5$ and $55~{\rm km}~{\rm s}^{-1}$, which would reduce $\Delta z_e$ by $\sim 30\%$. 

The results of our small-box simulations pose extra uncertainties due to cosmic variance and fluctuations in local density. C20 found the impact of the streaming on recombination rate can be several times smaller in larger simulation boxes or with a lower ionizing intensity (e.g., $\Gamma_{-12}=0.3$).   Our choice of the ionization background intensity, $\Gamma_{-12}=9.2$, is more relevant to the early phase of the reionization, where the HII regions exist only in the vicinity of ionization sources. As the HII regions expand to fill the entire space, the mean value should decrease, especially in the decelerating reionization scenario. The box-size test of C20 suggests that the velocity streaming effect tends to be smaller in larger boxes due to massive halos that are less sensitive to the streaming motion. Indeed, our simulation box lacks halos more massive than $10^7~M_\odot$. However, halos with $10^7\lesssim M \lesssim 10^9~M_\odot$ are known to form a certain amount of stars during the pre-reionization phase \citep[e.g.,][]{2019ApJ...878...98J}, violating one of the key assumptions of this study: no star-formation in the simulated volume. Thus, neglecting star-formation inside these halos would likely overestimate the ionization photon budget. We shall explore this possibility in our future study.

Considering the patchy nature of reionization, which our one-zone model does not account for, the fluctuations in the UV background and IGM temperature caused by the clustering of ionizing sources should also contribute significantly to the $\Delta z_e$ from quasar sightlines on top of the contribution from the streaming velocity fluctuations. Given that various factors mentioned above ($\Gamma_{-12}$, $T_{\rm gas}$, box size, the actual range of $V_{cb,i}$, and self-regulation of ionizing sources) can make $\Delta z_e$ of several times smaller than in our most extreme scenario ($\Delta z_e = 0.7$), it is more realistic to consider a value like $\Delta z_e\sim 0.1-0.2$, which would help to explain observed $\Delta z_e$ when combined with the other contributions.

The long wavelength of the streaming velocity fluctuations would also leave a unique signature on the reionization observable like the 21cm background, the near-infrared background, and the CMB via the kinetic Sunyaev$-$Zel'dovich effect. It is worth exploring these effects in the ionization field using seminumerical \citep[e.g.,][]{2011MNRAS.411..955M,2014MNRAS.440.1662S,2020arXiv200408401H} or fully numerical simulations of reionization \citep[e.g., CoDa simulation;][]{2016MNRAS.463.1462O,2020MNRAS.496.4087O}.

\section*{Acknowledgement}
The authors thank A. Mesinger,  C. Cain, A. D'Aloisio, and the  anonymous referee for helpful comments on this work. This work was supported by the World Premier International Research Center Initiative (WPI), MEXT, Japan. H.P. was supported by JSPS KAKENHI grant No. 19K23455. K.A. was supported by NRF-2016R1D1A1B04935414 and NRF-2016R1A5A1013277, and appreciates APCTP and IPMU for their hospitality during completion of this work. S.H. was supported by JSPS KAKENHI grant No. 18J01296. P.R.S. was supported in part by U.S. NSF grant No. AST-1009799, NASA grant No. NNX11AE09G, NASA/JPL grant No. RSA Nos. 1492788 and 1515294, and supercomputer resources from NSF XSEDE grant No. TG-AST090005 and the Texas Advanced Computing Center (TACC) at the University of Texas at Austin. Numerical computations were carried out on Cray XC50 and PC cluster at Center for Computational Astrophysics, National Astronomical Observatory of Japan.

\end{CJK}
\bibliographystyle{apj}
\bibliography{reference}

\begin{thebibliography}{}
\expandafter\ifx\csname natexlab\endcsname\relax\def\natexlab#1{#1}\fi

\bibitem[{{Ahn}(2016)}]{2016ApJ...830...68A}
{Ahn}, K. 2016, \apj, 830, 68

\bibitem[{{Ahn} {et~al.}(2012){Ahn}, {Iliev}, {Shapiro}, {Mellema}, {Koda}, \&
  {Mao}}]{2012ApJ...756L..16A}
{Ahn}, K., {Iliev}, I.~T., {Shapiro}, P.~R., {et~al.} 2012, \apjl, 756, L16

\bibitem[{{Ahn} \& {Smith}(2018)}]{2018ApJ...869...76A}
{Ahn}, K., \& {Smith}, B.~D. 2018, \apj, 869, 76

\bibitem[{{Asaba} {et~al.}(2016){Asaba}, {Ichiki}, \&
  {Tashiro}}]{2016PhRvD..93b3518A}
{Asaba}, S., {Ichiki}, K., \& {Tashiro}, H. 2016, \prd, 93, 023518

\bibitem[{{Ba{\~n}ados} {et~al.}(2018){Ba{\~n}ados}, {Venemans},
  {Mazzucchelli}, {Farina}, {Walter}, {Wang}, {Decarli}, {Stern}, {Fan},
  {Davies}, {Hennawi}, {Simcoe}, {Turner}, {Rix}, {Yang}, {Kelson}, {Rudie}, \&
  {Winters}}]{2018Natur.553..473B}
{Ba{\~n}ados}, E., {Venemans}, B.~P., {Mazzucchelli}, C., {et~al.} 2018, \nat,
  553, 473

\bibitem[{{Becker} {et~al.}(2015){Becker}, {Bolton}, {Madau}, {Pettini},
  {Ryan-Weber}, \& {Venemans}}]{2015MNRAS.447.3402B}
{Becker}, G.~D., {Bolton}, J.~S., {Madau}, P., {et~al.} 2015, \mnras, 447, 3402

\bibitem[{{Bolton} {et~al.}(2011){Bolton}, {Haehnelt}, {Warren}, {Hewett},
  {Mortlock}, {Venemans}, {McMahon}, \& {Simpson}}]{2011MNRAS.416L..70B}
{Bolton}, J.~S., {Haehnelt}, M.~G., {Warren}, S.~J., {et~al.} 2011, \mnras,
  416, L70

\bibitem[{{Bouwens} {et~al.}(2015){Bouwens}, {Illingworth}, {Oesch}, {Trenti},
  {Labb{\'e}}, {Bradley}, {Carollo}, {van Dokkum}, {Gonzalez}, {Holwerda},
  {Franx}, {Spitler}, {Smit}, \& {Magee}}]{2015ApJ...803...34B}
{Bouwens}, R.~J., {Illingworth}, G.~D., {Oesch}, P.~A., {et~al.} 2015, \apj,
  803, 34

\bibitem[{{Cain} {et~al.}(2020){Cain}, {D'Aloisio}, {Ir{\v{s}}i{\v{c}}},
  {McQuinn}, \& {Trac}}]{2020ApJ...898..168C}
{Cain}, C., {D'Aloisio}, A., {Ir{\v{s}}i{\v{c}}}, V., {McQuinn}, M., \& {Trac},
  H. 2020, \apj, 898, 168

\bibitem[{{Chardin} {et~al.}(2017){Chardin}, {Puchwein}, \&
  {Haehnelt}}]{2017MNRAS.465.3429C}
{Chardin}, J., {Puchwein}, E., \& {Haehnelt}, M.~G. 2017, \mnras, 465, 3429

\bibitem[{{D'Aloisio} {et~al.}(2018){D'Aloisio}, {McQuinn}, {Davies}, \&
  {Furlanetto}}]{2018MNRAS.473..560D}
{D'Aloisio}, A., {McQuinn}, M., {Davies}, F.~B., \& {Furlanetto}, S.~R. 2018,
  \mnras, 473, 560

\bibitem[{{D'Aloisio} {et~al.}(2015){D'Aloisio}, {McQuinn}, \&
  {Trac}}]{2015ApJ...813L..38D}
{D'Aloisio}, A., {McQuinn}, M., \& {Trac}, H. 2015, \apjl, 813, L38

\bibitem[{{D'Aloisio} {et~al.}(2020){D'Aloisio}, {McQuinn}, {Trac}, {Cain}, \&
  {Mesinger}}]{2020ApJ...898..149D}
{D'Aloisio}, A., {McQuinn}, M., {Trac}, H., {Cain}, C., \& {Mesinger}, A. 2020,
  \apj, 898, 149

\bibitem[{{D'Aloisio} {et~al.}(2017){D'Aloisio}, {Upton Sanderbeck}, {McQuinn},
  {Trac}, \& {Shapiro}}]{2017MNRAS.468.4691D}
{D'Aloisio}, A., {Upton Sanderbeck}, P.~R., {McQuinn}, M., {Trac}, H., \&
  {Shapiro}, P.~R. 2017, \mnras, 468, 4691

\bibitem[{{Davies} \& {Furlanetto}(2016)}]{2016MNRAS.460.1328D}
{Davies}, F.~B., \& {Furlanetto}, S.~R. 2016, \mnras, 460, 1328

\bibitem[{{Davies} {et~al.}(2018){Davies}, {Hennawi}, {Ba{\~n}ados},
  {Luki{\'c}}, {Decarli}, {Fan}, {Farina}, {Mazzucchelli}, {Rix}, {Venemans},
  {Walter}, {Wang}, \& {Yang}}]{2018ApJ...864..142D}
{Davies}, F.~B., {Hennawi}, J.~F., {Ba{\~n}ados}, E., {et~al.} 2018, \apj, 864,
  142

\bibitem[{{Eide} {et~al.}(2018){Eide}, {Graziani}, {Ciardi}, {Feng},
  {Kakiichi}, \& {Di Matteo}}]{2018MNRAS.476.1174E}
{Eide}, M.~B., {Graziani}, L., {Ciardi}, B., {et~al.} 2018, \mnras, 476, 1174

\bibitem[{{Emberson} {et~al.}(2013){Emberson}, {Thomas}, \&
  {Alvarez}}]{2013ApJ...763..146E}
{Emberson}, J.~D., {Thomas}, R.~M., \& {Alvarez}, M.~A. 2013, \apj, 763, 146

\bibitem[{{Fan} {et~al.}(2006){Fan}, {Carilli}, \&
  {Keating}}]{2006ARA&A..44..415F}
{Fan}, X., {Carilli}, C.~L., \& {Keating}, B. 2006, \araa, 44, 415

\bibitem[{{Fialkov} {et~al.}(2014){Fialkov}, {Barkana}, {Pinhas}, \&
  {Visbal}}]{2014MNRAS.437L..36F}
{Fialkov}, A., {Barkana}, R., {Pinhas}, A., \& {Visbal}, E. 2014, \mnras, 437,
  L36

\bibitem[{{Finkelstein} {et~al.}(2015){Finkelstein}, {Ryan}, {Papovich},
  {Dickinson}, {Song}, {Somerville}, {Ferguson}, {Salmon}, {Giavalisco},
  {Koekemoer}, {Ashby}, {Behroozi}, {Castellano}, {Dunlop}, {Faber}, {Fazio},
  {Fontana}, {Grogin}, {Hathi}, {Jaacks}, {Kocevski}, {Livermore}, {McLure},
  {Merlin}, {Mobasher}, {Newman}, {Rafelski}, {Tilvi}, \&
  {Willner}}]{2015ApJ...810...71F}
{Finkelstein}, S.~L., {Ryan}, Russell~E., J., {Papovich}, C., {et~al.} 2015,
  \apj, 810, 71

\bibitem[{{Finkelstein} {et~al.}(2019){Finkelstein}, {D'Aloisio},
  {Paardekooper}, {Ryan}, {Behroozi}, {Finlator}, {Livermore}, {Upton
  Sanderbeck}, {Dalla Vecchia}, \& {Khochfar}}]{2019ApJ...879...36F}
{Finkelstein}, S.~L., {D'Aloisio}, A., {Paardekooper}, J.-P., {et~al.} 2019,
  \apj, 879, 36

\bibitem[{{Fontana} {et~al.}(2010){Fontana}, {Vanzella}, {Pentericci},
  {Castellano}, {Giavalisco}, {Grazian}, {Boutsia}, {Cristiani}, {Dickinson},
  {Giallongo}, {Maiolino}, {Moorwood}, \& {Santini}}]{2010ApJ...725L.205F}
{Fontana}, A., {Vanzella}, E., {Pentericci}, L., {et~al.} 2010, \apjl, 725,
  L205

\bibitem[{{Gallerani} {et~al.}(2008){Gallerani}, {Ferrara}, {Fan}, \&
  {Choudhury}}]{2008MNRAS.386..359G}
{Gallerani}, S., {Ferrara}, A., {Fan}, X., \& {Choudhury}, T.~R. 2008, \mnras,
  386, 359

\bibitem[{{Greif} {et~al.}(2011){Greif}, {White}, {Klessen}, \&
  {Springel}}]{2011ApJ...736..147G}
{Greif}, T.~H., {White}, S. D.~M., {Klessen}, R.~S., \& {Springel}, V. 2011,
  \apj, 736, 147

\bibitem[{{Haiman} {et~al.}(2000){Haiman}, {Abel}, \&
  {Rees}}]{2000ApJ...534...11H}
{Haiman}, Z., {Abel}, T., \& {Rees}, M.~J. 2000, \apj, 534, 11

\bibitem[{{Hinshaw} {et~al.}(2013){Hinshaw}, {Larson}, {Komatsu}, {Spergel},
  {Bennett}, {Dunkley}, {Nolta}, {Halpern}, {Hill}, {Odegard}, {Page}, {Smith},
  {Weiland}, {Gold}, {Jarosik}, {Kogut}, {Limon}, {Meyer}, {Tucker}, {Wollack},
  \& {Wright}}]{2013ApJS..208...19H}
{Hinshaw}, G., {Larson}, D., {Komatsu}, E., {et~al.} 2013, \apjs, 208, 19

\bibitem[{{Hirano} {et~al.}(2017){Hirano}, {Hosokawa}, {Yoshida}, \&
  {Kuiper}}]{2017Sci...357.1375H}
{Hirano}, S., {Hosokawa}, T., {Yoshida}, N., \& {Kuiper}, R. 2017, Science,
  357, 1375

\bibitem[{{Hirata}(2018)}]{2018MNRAS.474.2173H}
{Hirata}, C.~M. 2018, \mnras, 474, 2173

\bibitem[{{Hoag} {et~al.}(2019){Hoag}, {Brada{\v{c}}}, {Huang}, {Mason},
  {Treu}, {Schmidt}, {Trenti}, {Strait}, {Lemaux}, {Finney}, \&
  {Paddock}}]{2019ApJ...878...12H}
{Hoag}, A., {Brada{\v{c}}}, M., {Huang}, K., {et~al.} 2019, \apj, 878, 12

\bibitem[{{Hutter} {et~al.}(2020{\natexlab{a}}){Hutter}, {Dayal}, {Legrand},
  {Gottl{\"o}ber}, \& {Yepes}}]{2020arXiv200813215H}
{Hutter}, A., {Dayal}, P., {Legrand}, L., {Gottl{\"o}ber}, S., \& {Yepes}, G.
  2020{\natexlab{a}}, arXiv e-prints, arXiv:2008.13215

\bibitem[{{Hutter} {et~al.}(2020{\natexlab{b}}){Hutter}, {Dayal}, {Yepes},
  {Gottl{\"o}ber}, {Legrand}, \& {Ucci}}]{2020arXiv200408401H}
{Hutter}, A., {Dayal}, P., {Yepes}, G., {et~al.} 2020{\natexlab{b}}, arXiv
  e-prints, arXiv:2004.08401

\bibitem[{{Iliev} {et~al.}(2007){Iliev}, {Mellema}, {Shapiro}, \&
  {Pen}}]{2007MNRAS.376..534I}
{Iliev}, I.~T., {Mellema}, G., {Shapiro}, P.~R., \& {Pen}, U.-L. 2007, \mnras,
  376, 534

\bibitem[{{Iliev} {et~al.}(2005{\natexlab{a}}){Iliev}, {Scannapieco}, \&
  {Shapiro}}]{2005ApJ...624..491I}
{Iliev}, I.~T., {Scannapieco}, E., \& {Shapiro}, P.~R. 2005{\natexlab{a}},
  \apj, 624, 491

\bibitem[{{Iliev} {et~al.}(2005{\natexlab{b}}){Iliev}, {Shapiro}, \&
  {Raga}}]{2005MNRAS.361..405I}
{Iliev}, I.~T., {Shapiro}, P.~R., \& {Raga}, A.~C. 2005{\natexlab{b}}, \mnras,
  361, 405

\bibitem[{{Inoue} {et~al.}(2018){Inoue}, {Hasegawa}, {Ishiyama}, {Yajima},
  {Shimizu}, {Umemura}, {Konno}, {Harikane}, {Shibuya}, {Ouchi}, {Shimasaku},
  {Ono}, {Kusakabe}, {Higuchi}, \& {Lee}}]{2018PASJ...70...55I}
{Inoue}, A.~K., {Hasegawa}, K., {Ishiyama}, T., {et~al.} 2018, \pasj, 70, 55

\bibitem[{{Jeon} {et~al.}(2019){Jeon}, {Besla}, \&
  {Bromm}}]{2019ApJ...878...98J}
{Jeon}, M., {Besla}, G., \& {Bromm}, V. 2019, \apj, 878, 98

\bibitem[{{Jung} {et~al.}(2018){Jung}, {Finkelstein}, {Livermore}, {Dickinson},
  {Larson}, {Papovich}, {Song}, {Tilvi}, \& {Wold}}]{2018ApJ...864..103J}
{Jung}, I., {Finkelstein}, S.~L., {Livermore}, R.~C., {et~al.} 2018, \apj, 864,
  103

\bibitem[{{Jung} {et~al.}(2020){Jung}, {Finkelstein}, {Dickinson}, {Hutchison},
  {Larson}, {Papovich}, {Pentericci}, {Straughn}, {Guo}, {Malhotra}, {Rhoads},
  {Song}, {Tilvi}, \& {Wold}}]{2020ApJ...904..144J}
{Jung}, I., {Finkelstein}, S.~L., {Dickinson}, M., {et~al.} 2020, \apj, 904,
  144

\bibitem[{{Keating} {et~al.}(2020){Keating}, {Weinberger}, {Kulkarni},
  {Haehnelt}, {Chardin}, \& {Aubert}}]{2020MNRAS.491.1736K}
{Keating}, L.~C., {Weinberger}, L.~H., {Kulkarni}, G., {et~al.} 2020, \mnras,
  491, 1736

\bibitem[{{Konno} {et~al.}(2018){Konno}, {Ouchi}, {Shibuya}, {Ono},
  {Shimasaku}, {Taniguchi}, {Nagao}, {Kobayashi}, {Kajisawa}, {Kashikawa},
  {Inoue}, {Oguri}, {Furusawa}, {Goto}, {Harikane}, {Higuchi}, {Komiyama},
  {Kusakabe}, {Miyazaki}, {Nakajima}, \& {Wang}}]{2018PASJ...70S..16K}
{Konno}, A., {Ouchi}, M., {Shibuya}, T., {et~al.} 2018, \pasj, 70, S16

\bibitem[{{Kulkarni} {et~al.}(2019){Kulkarni}, {Keating}, {Haehnelt}, {Bosman},
  {Puchwein}, {Chardin}, \& {Aubert}}]{2019MNRAS.485L..24K}
{Kulkarni}, G., {Keating}, L.~C., {Haehnelt}, M.~G., {et~al.} 2019, \mnras,
  485, L24

\bibitem[{{Madau} {et~al.}(1999){Madau}, {Haardt}, \&
  {Rees}}]{1999ApJ...514..648M}
{Madau}, P., {Haardt}, F., \& {Rees}, M.~J. 1999, \apj, 514, 648

\bibitem[{{Mao} {et~al.}(2020){Mao}, {Koda}, {Shapiro}, {Iliev}, {Mellema},
  {Park}, {Ahn}, \& {Bianco}}]{2020MNRAS.491.1600M}
{Mao}, Y., {Koda}, J., {Shapiro}, P.~R., {et~al.} 2020, \mnras, 491, 1600

\bibitem[{{Mason} {et~al.}(2018){Mason}, {Treu}, {Dijkstra}, {Mesinger},
  {Trenti}, {Pentericci}, {de Barros}, \& {Vanzella}}]{2018ApJ...856....2M}
{Mason}, C.~A., {Treu}, T., {Dijkstra}, M., {et~al.} 2018, \apj, 856, 2

\bibitem[{{McGreer} {et~al.}(2015){McGreer}, {Mesinger}, \&
  {D'Odorico}}]{2015MNRAS.447..499M}
{McGreer}, I.~D., {Mesinger}, A., \& {D'Odorico}, V. 2015, \mnras, 447, 499

\bibitem[{{McLure} {et~al.}(2013){McLure}, {Dunlop}, {Bowler}, {Curtis-Lake},
  {Schenker}, {Ellis}, {Robertson}, {Koekemoer}, {Rogers}, {Ono}, {Ouchi},
  {Charlot}, {Wild}, {Stark}, {Furlanetto}, {Cirasuolo}, \&
  {Targett}}]{2013MNRAS.432.2696M}
{McLure}, R.~J., {Dunlop}, J.~S., {Bowler}, R.~A.~A., {et~al.} 2013, \mnras,
  432, 2696

\bibitem[{{McQuinn} \& {O'Leary}(2012)}]{2012ApJ...760....3M}
{McQuinn}, M., \& {O'Leary}, R.~M. 2012, \apj, 760, 3

\bibitem[{{Mesinger} {et~al.}(2011){Mesinger}, {Furlanetto}, \&
  {Cen}}]{2011MNRAS.411..955M}
{Mesinger}, A., {Furlanetto}, S., \& {Cen}, R. 2011, \mnras, 411, 955

\bibitem[{{Montero-Camacho} {et~al.}(2019){Montero-Camacho}, {Hirata},
  {Martini}, \& {Honscheid}}]{2019MNRAS.487.1047M}
{Montero-Camacho}, P., {Hirata}, C.~M., {Martini}, P., \& {Honscheid}, K. 2019,
  \mnras, 487, 1047

\bibitem[{{Montero-Camacho} \& {Mao}(2020)}]{2020MNRAS.499.1640M}
{Montero-Camacho}, P., \& {Mao}, Y. 2020, \mnras, 499, 1640

\bibitem[{{Mortlock} {et~al.}(2011){Mortlock}, {Warren}, {Venemans}, {Patel},
  {Hewett}, {McMahon}, {Simpson}, {Theuns}, {Gonz{\'a}les-Solares}, {Adamson},
  {Dye}, {Hambly}, {Hirst}, {Irwin}, {Kuiper}, {Lawrence}, \&
  {R{\"o}ttgering}}]{2011Natur.474..616M}
{Mortlock}, D.~J., {Warren}, S.~J., {Venemans}, B.~P., {et~al.} 2011, \nat,
  474, 616

\bibitem[{{Mu{\~n}oz}(2019)}]{2019PhRvD.100f3538M}
{Mu{\~n}oz}, J.~B. 2019, \prd, 100, 063538

\bibitem[{{Naoz} {et~al.}(2013){Naoz}, {Yoshida}, \&
  {Gnedin}}]{2013ApJ...763...27N}
{Naoz}, S., {Yoshida}, N., \& {Gnedin}, N.~Y. 2013, \apj, 763, 27

\bibitem[{{Nasir} \& {D'Aloisio}(2020)}]{2020MNRAS.494.3080N}
{Nasir}, F., \& {D'Aloisio}, A. 2020, \mnras, 494, 3080

\bibitem[{{Ocvirk} {et~al.}(2016){Ocvirk}, {Gillet}, {Shapiro}, {Aubert},
  {Iliev}, {Teyssier}, {Yepes}, {Choi}, {Sullivan}, {Knebe}, {Gottl{\"o}ber},
  {D'Aloisio}, {Park}, {Hoffman}, \& {Stranex}}]{2016MNRAS.463.1462O}
{Ocvirk}, P., {Gillet}, N., {Shapiro}, P.~R., {et~al.} 2016, \mnras, 463, 1462

\bibitem[{{Ocvirk} {et~al.}(2020){Ocvirk}, {Aubert}, {Sorce}, {Shapiro},
  {Deparis}, {Dawoodbhoy}, {Lewis}, {Teyssier}, {Yepes}, {Gottl{\"o}ber},
  {Ahn}, {Iliev}, \& {Hoffman}}]{2020MNRAS.496.4087O}
{Ocvirk}, P., {Aubert}, D., {Sorce}, J.~G., {et~al.} 2020, \mnras, 496, 4087

\bibitem[{{O'Leary} \& {McQuinn}(2012)}]{2012ApJ...760....4O}
{O'Leary}, R.~M., \& {McQuinn}, M. 2012, \apj, 760, 4

\bibitem[{{Park} {et~al.}(2020){Park}, {Ahn}, {Yoshida}, \&
  {Hirano}}]{2020ApJ...900...30P}
{Park}, H., {Ahn}, K., {Yoshida}, N., \& {Hirano}, S. 2020, \apj, 900, 30

\bibitem[{{Park} {et~al.}(2016){Park}, {Shapiro}, {Choi}, {Yoshida}, {Hirano},
  \& {Ahn}}]{2016ApJ...831...86P}
{Park}, H., {Shapiro}, P.~R., {Choi}, J.-h., {et~al.} 2016, \apj, 831, 86

\bibitem[{{Pawlik} {et~al.}(2009){Pawlik}, {Schaye}, \& {van
  Scherpenzeel}}]{2009MNRAS.394.1812P}
{Pawlik}, A.~H., {Schaye}, J., \& {van Scherpenzeel}, E. 2009, \mnras, 394,
  1812

\bibitem[{{Pentericci} {et~al.}(2011){Pentericci}, {Fontana}, {Vanzella},
  {Castellano}, {Grazian}, {Dijkstra}, {Boutsia}, {Cristiani}, {Dickinson},
  {Giallongo}, {Giavalisco}, {Maiolino}, {Moorwood}, {Paris}, \&
  {Santini}}]{2011ApJ...743..132P}
{Pentericci}, L., {Fontana}, A., {Vanzella}, E., {et~al.} 2011, \apj, 743, 132

\bibitem[{{Planck Collaboration} {et~al.}(2018){Planck Collaboration},
  {Aghanim}, {Akrami}, {Ashdown}, {Aumont}, {Baccigalupi}, {Ballardini},
  {Banday}, {Barreiro}, {Bartolo}, {Basak}, {Battye}, {Benabed}, {Bernard},
  {Bersanelli}, {Bielewicz}, {Bock}, {Bond}, {Borrill}, {Bouchet}, {Boulanger},
  {Bucher}, {Burigana}, {Butler}, {Calabrese}, {Cardoso}, {Carron},
  {Challinor}, {Chiang}, {Chluba}, {Colombo}, {Combet}, {Contreras}, {Crill},
  {Cuttaia}, {de Bernardis}, {de Zotti}, {Delabrouille}, {Delouis}, {Di
  Valentino}, {Diego}, {Dor{\'e}}, {Douspis}, {Ducout}, {Dupac}, {Dusini},
  {Efstathiou}, {Elsner}, {En{\ss}lin}, {Eriksen}, {Fantaye}, {Farhang},
  {Fergusson}, {Fernandez-Cobos}, {Finelli}, {Forastieri}, {Frailis},
  {Fraisse}, {Franceschi}, {Frolov}, {Galeotta}, {Galli}, {Ganga},
  {G{\'e}nova-Santos}, {Gerbino}, {Ghosh}, {Gonz{\'a}lez-Nuevo}, {G{\'o}rski},
  {Gratton}, {Gruppuso}, {Gudmundsson}, {Hamann}, {Handley}, {Hansen},
  {Herranz}, {Hildebrandt}, {Hivon}, {Huang}, {Jaffe}, {Jones}, {Karakci},
  {Keih{\"a}nen}, {Keskitalo}, {Kiiveri}, {Kim}, {Kisner}, {Knox},
  {Krachmalnicoff}, {Kunz}, {Kurki-Suonio}, {Lagache}, {Lamarre}, {Lasenby},
  {Lattanzi}, {Lawrence}, {Le Jeune}, {Lemos}, {Lesgourgues}, {Levrier},
  {Lewis}, {Liguori}, {Lilje}, {Lilley}, {Lindholm}, {L{\'o}pez-Caniego},
  {Lubin}, {Ma}, {Mac{\'\i}as-P{\'e}rez}, {Maggio}, {Maino}, {Mandolesi},
  {Mangilli}, {Marcos-Caballero}, {Maris}, {Martin}, {Martinelli},
  {Mart{\'\i}nez-Gonz{\'a}lez}, {Matarrese}, {Mauri}, {McEwen}, {Meinhold},
  {Melchiorri}, {Mennella}, {Migliaccio}, {Millea}, {Mitra},
  {Miville-Desch{\^e}nes}, {Molinari}, {Montier}, {Morgante}, {Moss}, {Natoli},
  {N{\o}rgaard-Nielsen}, {Pagano}, {Paoletti}, {Partridge}, {Patanchon},
  {Peiris}, {Perrotta}, {Pettorino}, {Piacentini}, {Polastri}, {Polenta},
  {Puget}, {Rachen}, {Reinecke}, {Remazeilles}, {Renzi}, {Rocha}, {Rosset},
  {Roudier}, {Rubi{\~n}o-Mart{\'\i}n}, {Ruiz-Granados}, {Salvati}, {Sandri},
  {Savelainen}, {Scott}, {Shellard}, {Sirignano}, {Sirri}, {Spencer},
  {Sunyaev}, {Suur-Uski}, {Tauber}, {Tavagnacco}, {Tenti}, {Toffolatti},
  {Tomasi}, {Trombetti}, {Valenziano}, {Valiviita}, {Van Tent}, {Vibert},
  {Vielva}, {Villa}, {Vittorio}, {Wand elt}, {Wehus}, {White}, {White},
  {Zacchei}, \& {Zonca}}]{2018arXiv180706209P}
{Planck Collaboration}, {Aghanim}, N., {Akrami}, Y., {et~al.} 2018, arXiv
  e-prints, arXiv:1807.06209

\bibitem[{{Richardson} {et~al.}(2013){Richardson}, {Scannapieco}, \&
  {Thacker}}]{2013ApJ...771...81R}
{Richardson}, M. L.~A., {Scannapieco}, E., \& {Thacker}, R.~J. 2013, \apj, 771,
  81

\bibitem[{{Ricotti} \& {Ostriker}(2004)}]{2004MNRAS.352..547R}
{Ricotti}, M., \& {Ostriker}, J.~P. 2004, \mnras, 352, 547

\bibitem[{{Robertson} {et~al.}(2015){Robertson}, {Ellis}, {Furlanetto}, \&
  {Dunlop}}]{2015ApJ...802L..19R}
{Robertson}, B.~E., {Ellis}, R.~S., {Furlanetto}, S.~R., \& {Dunlop}, J.~S.
  2015, \apjl, 802, L19

\bibitem[{{Robertson} {et~al.}(2013){Robertson}, {Furlanetto}, {Schneider},
  {Charlot}, {Ellis}, {Stark}, {McLure}, {Dunlop}, {Koekemoer}, {Schenker},
  {Ouchi}, {Ono}, {Curtis-Lake}, {Rogers}, {Bowler}, \&
  {Cirasuolo}}]{2013ApJ...768...71R}
{Robertson}, B.~E., {Furlanetto}, S.~R., {Schneider}, E., {et~al.} 2013, \apj,
  768, 71

\bibitem[{{Schauer} {et~al.}(2019){Schauer}, {Glover}, {Klessen}, \&
  {Ceverino}}]{2019MNRAS.484.3510S}
{Schauer}, A. T.~P., {Glover}, S. C.~O., {Klessen}, R.~S., \& {Ceverino}, D.
  2019, \mnras, 484, 3510

\bibitem[{{Schenker} {et~al.}(2012){Schenker}, {Stark}, {Ellis}, {Robertson},
  {Dunlop}, {McLure}, {Kneib}, \& {Richard}}]{2012ApJ...744..179S}
{Schenker}, M.~A., {Stark}, D.~P., {Ellis}, R.~S., {et~al.} 2012, \apj, 744,
  179

\bibitem[{{Schmidt} {et~al.}(2016){Schmidt}, {Treu}, {Brada{\v{c}}}, {Vulcani},
  {Huang}, {Hoag}, {Maseda}, {Guaita}, {Pentericci}, {Brammer}, {Dijkstra},
  {Dressler}, {Fontana}, {Henry}, {Jones}, {Mason}, {Trenti}, \&
  {Wang}}]{2016ApJ...818...38S}
{Schmidt}, K.~B., {Treu}, T., {Brada{\v{c}}}, M., {et~al.} 2016, \apj, 818, 38

\bibitem[{{Shapiro} \& {Giroux}(1987)}]{1987ApJ...321L.107S}
{Shapiro}, P.~R., \& {Giroux}, M.~L. 1987, \apjl, 321, L107

\bibitem[{{Shapiro} {et~al.}(2004){Shapiro}, {Iliev}, \&
  {Raga}}]{2004MNRAS.348..753S}
{Shapiro}, P.~R., {Iliev}, I.~T., \& {Raga}, A.~C. 2004, \mnras, 348, 753

\bibitem[{{Shapiro} {et~al.}(2003){Shapiro}, {Iliev}, {Raga}, \&
  {Martel}}]{2003AIPC..666...89S}
{Shapiro}, P.~R., {Iliev}, I.~T., {Raga}, A.~C., \& {Martel}, H. 2003, in
  American Institute of Physics Conference Series, Vol. 666, The Emergence of
  Cosmic Structure, ed. S.~H. {Holt} \& C.~S. {Reynolds}, 89--92

\bibitem[{{Sobacchi} \& {Mesinger}(2014)}]{2014MNRAS.440.1662S}
{Sobacchi}, E., \& {Mesinger}, A. 2014, \mnras, 440, 1662

\bibitem[{{Songaila}(2004)}]{2004AJ....127.2598S}
{Songaila}, A. 2004, \aj, 127, 2598

\bibitem[{{Springel}(2005)}]{2005MNRAS.364.1105S}
{Springel}, V. 2005, \mnras, 364, 1105

\bibitem[{{Springel} {et~al.}(2001){Springel}, {Yoshida}, \&
  {White}}]{2001NewA....6...79S}
{Springel}, V., {Yoshida}, N., \& {White}, S. D.~M. 2001, \na, 6, 79

\bibitem[{{Stacy} {et~al.}(2011){Stacy}, {Bromm}, \&
  {Loeb}}]{2011ApJ...730L...1S}
{Stacy}, A., {Bromm}, V., \& {Loeb}, A. 2011, \apjl, 730, L1

\bibitem[{{Tseliakhovich} \& {Hirata}(2010)}]{2010PhRvD..82h3520T}
{Tseliakhovich}, D., \& {Hirata}, C. 2010, \prd, 82, 083520

\bibitem[{{Visbal} {et~al.}(2012){Visbal}, {Barkana}, {Fialkov},
  {Tseliakhovich}, \& {Hirata}}]{2012Natur.487...70V}
{Visbal}, E., {Barkana}, R., {Fialkov}, A., {Tseliakhovich}, D., \& {Hirata},
  C.~M. 2012, \nat, 487, 70

\bibitem[{{Whitler} {et~al.}(2020){Whitler}, {Mason}, {Ren}, {Dijkstra},
  {Mesinger}, {Pentericci}, {Trenti}, \& {Treu}}]{2020MNRAS.495.3602W}
{Whitler}, L.~R., {Mason}, C.~A., {Ren}, K., {et~al.} 2020, \mnras, 495, 3602

\bibitem[{{Yoshida} {et~al.}(2007){Yoshida}, {Oh}, {Kitayama}, \&
  {Hernquist}}]{2007ApJ...663..687Y}
{Yoshida}, N., {Oh}, S.~P., {Kitayama}, T., \& {Hernquist}, L. 2007, \apj, 663,
  687

\bibitem[{{Yoshida} {et~al.}(2006){Yoshida}, {Omukai}, {Hernquist}, \&
  {Abel}}]{2006ApJ...652....6Y}
{Yoshida}, N., {Omukai}, K., {Hernquist}, L., \& {Abel}, T. 2006, \apj, 652, 6

\end{thebibliography}
\end{document}